\documentclass[aps,prd,nofootinbib,showpacs,superscriptaddress,twocolumn,floatfix]{revtex4-1}

\usepackage{amsmath,amssymb}
\usepackage[utf8]{inputenc}
\usepackage{graphicx}
\usepackage{mathrsfs}
\usepackage{bm}
\usepackage{bbm}
\usepackage{indentfirst}
\usepackage{epstopdf}
\usepackage{color}
\usepackage{xcolor}
\usepackage{amssymb, amsmath,color, hyperref, graphicx}
\usepackage[mathlines]{lineno}

\usepackage{braket}
\usepackage{placeins}
\usepackage{multirow}
\usepackage{slashed}
\usepackage{physics}

\newcommand{\chiBQ}{$\chi^{\rm BQ}_{11}$}
\newcommand{\chiBS}{$\chi^{\rm BS}_{11}$}
\newcommand{\chiQS}{$\chi^{\rm QS}_{11}$}
\newcommand{\chitB}{$\chi^{\rm B}_{2}$}
\newcommand{\chitQ}{$\chi^{\rm Q}_{2}$}
\newcommand{\chitS}{$\chi^{\rm S}_{2}$}

\begin{document}

\title{Second order fluctuations of conserved charges in external magnetic fields}

\author{Heng-Tong Ding}
\address{Key Laboratory of Quark and Lepton Physics (MOE) and Institute of
Particle Physics, Central China Normal University, Wuhan 430079, China}
\author{Jin-Biao Gu}
\affiliation{Key Laboratory of Quark and Lepton Physics (MOE) and Institute of
Particle Physics, Central China Normal University, Wuhan 430079, China}
\author{Arpith Kumar}
\affiliation{Key Laboratory of Quark and Lepton Physics (MOE) and Institute of
Particle Physics, Central China Normal University, Wuhan 430079, China}
\author{Sheng-Tai Li}
\affiliation{Key Laboratory of Quark and Lepton Physics (MOE) and Institute of
Particle Physics, Central China Normal University, Wuhan 430079, China}


\date{\today}
\begin{abstract}
We present a first-principles lattice QCD investigation of second-order fluctuations of and correlations among conserved charges---baryon number (B), electric charge (Q), and strangeness (S)---in the presence of external magnetic fields. Our study employs lattice simulations of (2+1)-flavor QCD with physical pion masses using highly improved staggered fermions on $48^3\times12$ and $32^3\times8$ lattices, covering a wide range of magnetic field strengths up to $eB \simeq 0.8$ GeV$^2$. 
We identify clear signals of magnetic field-induced modifications to these fluctuations and correlations, with the baryon-electric charge correlation, $\chi^{\rm BQ}_{11}$, exhibiting particularly strong sensitivity to the magnetic field. To bridge theoretical predictions with experimental observables, we implement systematic kinematic cuts that emulate detector acceptances of the STAR and ALICE experiments within the hadron resonance gas (HRG) model and construct proxy observables for fluctuations measurable in heavy-ion collision experiments. Our findings highlight $\chi^{\rm BQ}_{11}$ as a promising ``magnetometer" for probing the presence of magnetic fields in QCD matter. 
Furthermore, we explore experimentally relevant ratios involving $\chi^{\rm BQ}_{11}$, demonstrating their potential in mitigating volume effects and enhancing sensitivity to magnetic fields in collision environments. Additionally, we assess the limitations of the HRG model at strong magnetic fields, revealing deviations that indicate nontrivial modifications to hadronic degrees of freedom. These results offer new insights into the interplay between thermal and magnetic effects in the QCD phase diagram and provide experimentally relevant guidance for the detection of magnetic fields in heavy-ion collisions.

\end{abstract}


\maketitle


\section{Introduction} 
\label{sec:intro}

Strong magnetic fields---attaining magnitudes comparable to the intrinsic scales of interactions---play a fundamental role in emergent phenomena spanning three cornerstone domains of modern physics: cosmology, astrophysics, and high energy nuclear physics. Primordial magnetic fields are conjectured to emerge from cosmological density perturbations during the electroweak phase transition, potentially shaping the evolution of the early Universe \cite{Vachaspati:1991nm,Enqvist:1993np,Baym:1995fk,Grasso:2000wj}. In astrophysics, strong field strengths significantly influence the integral mass-radii properties of compact stars, especially strongly magnetized neutron stars, so-called magnetars \cite{Duncan:1992hi,Harding:2006qn}. More recently, in high energy nuclear physics, off-central heavy-ion collisions at terrestrial laboratories have sparked a great deal of interest, as highly energetic and electrically charged spectator particles are expected to generate one of the strongest magnetic fields to exist---comparable to the quantum chromodynamics (QCD) scale $\Lambda_{\rm QCD}^2 $. Model estimates suggest strengths $eB \sim 5 \,M_{\pi}^2$ at the Relativistic Heavy Ion Collider and $eB \sim 70 \,M_{\pi}^2$ at the Large Hadron Collider for $Pb$/$Au$ collisions \cite{Deng:2012pc, Skokov:2009qp}. Although transient in nature, if the lifetime is sufficiently sustained by electrical conductivity and magnetism of the collision medium \cite{Astrakhantsev:2019zkr,Bali:2020bcn,Ding:2016hua}, such strong magnetic fields are expected to induce significant nonperturbative effects on the thermal and transport characteristics of the produced QCD matter \cite{Kharzeev:2012ph}.

Consequently, the nontrivial topology of QCD can manifest as striking macroscopic phenomena, most notably the chiral magnetic effect \cite{Kharzeev:2007jp,Kharzeev:2020jxw}. In the context of heavy-ion collisions, leveraging the accessibility of strong magnetic fields, it was first proposed in 2007 to explain the chiral anomaly and spatial charge separation from the topologically invariant nature of gauge field configurations. The phenomenologically relevant quest to observe these field-induced effects and understand their origin has garnered intensive ongoing experimental and theoretical studies \cite{Cao:2021rwx,Endrodi:2024cqn,Adhikari:2024bfa}.

From a theoretical standpoint, interest emerged through complementary efforts involving low-energy effective theories and lattice QCD simulations. Within the framework of low-energy effective QCD models, semianalytical investigations on the strong-interaction matter in nonzero magnetic fields have been carried out using the linear sigma model \cite{Mizher:2010zb,Ferrari:2012yw,Ayala:2014gwa,Tawfik:2014hwa,Ayala:2020muk,Tawfik:2021eeb,Ayala:2023llp}, Nambu–Jona-Lasinio (NJL) model (along with their various extensions) \cite{Fukushima:2010fe,Skokov:2011ib,Fu:2013ica,Miransky:2015ava,Ayala:2016bbi,Farias:2016gmy,Cao:2021rwx,Sheng:2022ssp,Chahal:2023khc,Mao:2024gox,Ali:2024mnn},  hadron resonance gas (HRG) model \cite{Endrodi:2013cs,Bhattacharyya:2015pra,Fukushima:2016vix,Ferreira:2018pux,Kadam:2019rzo,Marczenko:2024kko,Vovchenko:2024wbg}, holographic QCD \cite{Rebhan:2009vc,Preis:2010cq,Fukushima:2013zga,Dudal:2018rki,Fukushima:2021got,Zhu:2023aaq} as well as AdS/CFT~\cite{Critelli:2014kra,Dudal:2014jfa,Yin:2021zhs}. Although these frameworks help us understand several key features of QCD dynamics, a comprehensive treatment of its nonperturbative aspects is generally achieved only through extensive numerical simulations, for which lattice regularization provides a robust, first-principles framework. Fortunately, contrary to an electric field or finite baryon density, introducing a magnetic background in the lattice QCD simulations does not give rise to numerical difficulties such as the infamous sign problem \cite{DElia:2010abb,Endrodi:2024cqn}.
Over the past decade, lattice QCD has substantially enhanced our understanding of strong magnetic fields effects, revealing significant modifications in key QCD properties, including thermodynamics~\cite{Bali:2014kia}, the phase diagram~\cite{Bali:2011qj,Ding:2020inp}, transport characteristics~\cite{Astrakhantsev:2019zkr,Velasco:2022gaw,Almirante:2024lqn,Brandt:2024fpc}, and in-medium hadron properties~\cite{Bonati:2015dka,Endrodi:2019whh,Ding:2022tqn,Ding:2025pbu}. Among the notable findings are now well-known phenomena such as a reduction in the transition temperature and inverse magnetic catalysis \cite{Bali:2011qj,Bali:2012zg,Bruckmann:2013oba,Bali:2013esa,Bonati:2016kxj,DElia:2018xwo,Endrodi:2019zrl,DElia:2021yvk,Brandt:2023dir}. Despite these purely theoretical advances, much remains to be understood about how external magnetic fields phenomenologically affect the underlying degrees of freedom in QCD. Notably, most lattice QCD studies in magnetic background thus far have centered on computing the chiral condensate~\cite{DElia:2011koc,Bali:2012zg,DElia:2018xwo,Endrodi:2019zrl,Ding:2020hxw,Ding:2025pbu,Ding:2022tqn}, a theoretical quantity that unfortunately cannot be directly measured in experiments.

Fluctuations of and correlations among net baryon number ($\rm B$), electric charge ($\rm Q$), and strangeness ($\rm S$) have long been recognized---both theoretically and experimentally---as powerful tools for probing changes in the degrees of freedom and the QCD phase structure in the absence of magnetic fields \cite{Asakawa:2000wh,Gavai:2001ie,Cheng:2008zh,Borsanyi:2011sw,HotQCD:2012fhj,Bollweg:2022rps,Bollweg:2024epj,Karsch:2010ck,Fu:2010ay,Fu:2018swz,Luo:2017faz,Pandav:2022xxx,Rustamov:2022hdi}. However, detailed studies of these fluctuations under magnetic fields remain scarce, with most existing work confined to effective models such as the HRG model~\cite{Fukushima:2016vix,Ferreira:2018pux,Bhattacharyya:2015pra,Kadam:2019rzo}, the Polyakov-NJL model \cite{Fu:2013ica,Mao:2024gox} and the Polyakov loop extended chiral SU(3) quark mean field model \cite{Chahal:2023khc}. Crucially, it is worth emphasizing that first-principles lattice simulations are indispensable for establishing model-independent benchmarks. Recently, with the very first work in 2021 \cite{Ding:2021cwv}, lattice studies of conserved changes in a magnetic background have begun to unfold. Although conducted employing a larger-than-physical pion mass, $M_{\pi} \simeq 220~\rm{MeV}$ at one single lattice cutoff, it paved interest for future studies. Later, the authors extended the work to a more physical setting with $M_{\pi} \simeq 135~\rm{MeV}$, and proposed the baryon-electric charge correlation as the magnetometer of QCD, which could be useful for probing the existence of magnetic fields in collision experiments \cite{Ding:2023bft,Ding:2025siy,Ding:2022uwj}. These works were followed by further investigations at the physical point in $(2+1+1)$-flavor QCD on finite lattices \cite{Borsanyi:2023buy,Borsanyi:2025mrf}, as well as fairly recent studies of the QCD equation of state \cite{Astrakhantsev:2024mat,Kumar:2025ikm}. 

In this work, we present an extensive (2+1)-flavor lattice QCD investigation for fluctuations of and correlations among conserved charges in the presence of strong magnetic fields with a physical pion mass, building upon our previous findings in Refs.~\cite{Ding:2023bft,Ding:2022uwj}.
Extending up to extremely strong field strengths $eB\simeq 0.8~{\rm GeV}^2 \sim 45 \,M_{\pi}^2$, we provide new insights into phenomenological relevant potential experimental signatures in collision experiments as well as the theoretically relevant intricate interplay between thermal and magnetic effects on QCD. Furthermore, we offer physical interpretations of our lattice QCD results using the HRG model at low temperatures, noticing a breakdown beyond a threshold field strength. To address experimental feasibility, we propose relevant proxies based on final detectable particles and implement systematic kinematic cuts on transverse momentum and pseudorapidity to account for detection challenges. At high temperatures, we explore how the magnetized ideal gas serves as a theoretical reference under strong magnetic fields.

The article is organized as follows. At the beginning of Sec. \ref{sec:fluc_HRG_ideal}, we introduce the fundamental definitions of generalized susceptibilities and establish the connection between the quark flavor basis and the physical conserved charge basis. We then outline a procedure to implement kinematic cuts on the proxies relevant to potential analyses within the framework of the magnetized HRG model for the ALICE and STAR experiments. Additionally, we briefly outline the ideal gas limit in a magnetized medium.  Section \ref{sec:lattice_setup} is devoted to details on the simulation setup and relevant physical parameters. Before investigating fluctuations and correlations, in Sec. \ref{sec:Tpc-eB} we first determine the crossover transition temperature in the presence of a magnetic field, $T_{pc}(eB)$, which reveals two distinct regimes: a relatively weak-field region and an extremely strong-field region.  In Sec. \ref{sec:weak-eB}, we present lattice results for leading-order susceptibilities in the presence of a relatively weak magnetic field. Subsequently, in Sec. \ref{sec:strong-eB}, we examine the case of extremely strong magnetic fields. Furthermore, we also discuss magnetic field-induced isospin-breaking effects in Sec. \ref{sec:isospin}.  Finally, in Sec. \ref{sec:summary}, we summarize our findings and present our conclusions. Additionally, we also list the parameters and statistics for the lattice QCD simulations in Appendix~\ref{app:stat} and provide supplementary materials for implementation of kinematic cuts on the proxy and plots to double ratios of leading order susceptibilities in Appendix~\ref{app:proxy_cuts}.

\section{Fluctuations and correlations of conserved charges in external magnetic fields}
\label{sec:fluc_HRG_ideal}

In the presence of an external magnetic field, the pressure $P$ in a thermal medium is a function of the temperature $T$, the magnetic field $B$ and a set of chemical potentials $(\mu_{\rm B}, \mu_{\rm Q}, \mu_{\rm S})$ corresponding to the conserved charges for net baryon number (${\rm B}$), electric charge (${\rm Q}$) and strangeness (${\rm S}$). The pressure can be expressed in terms of the logarithm of the grand canonical partition function $\mathcal{Z}$ as
\begin{equation}
    \frac{P}{T^4}=\frac{1}{V T^3} \ln \mathcal{Z}\left(V, T, \mu_{\rm B}, \mu_{\rm Q}, \mu_{\rm S},  B\right) .
\end{equation}
The chemical potentials of conserved charges can be rewritten in terms of the quark chemical potentials $(\mu_{u}, \mu_{d}, \mu_{s})$,
\begin{equation}
    \begin{aligned}
    & \mu_u=\frac{1}{3} \mu_{\mathrm{B}}+\frac{2}{3} \mu_{\mathrm{Q}}, \\
    & \mu_d=\frac{1}{3} \mu_{\mathrm{B}}-\frac{1}{3} \mu_{\mathrm{Q}}, \\
    & \mu_s=\frac{1}{3} \mu_{\mathrm{B}}-\frac{1}{3} \mu_{\mathrm{Q}}-\mu_{\mathrm{S}}.
    \end{aligned}
    \label{eq:BQS_uds_relation}
\end{equation}

Fluctuations of and correlations among conserved charges and quark numbers can be determined from lattice QCD calculations by taking derivatives of the pressure with respect to chemical potentials, evaluated at vanishing chemical potentials \cite{Allton:2002zi,Gavai:2003mf},
\begin{align}
    \chi_{i j k}^{{uds}}=\left.\frac{\partial^{i+j+k} P / T^4}{\partial\hat{\mu}_{u} ^i \partial\hat{\mu}_{d} ^j \partial\hat{\mu}_{s} ^k}\right|_{\hat{\mu}_{u,d,s}=0},\\
    \chi_{i j k}^{\mathrm{BQS}}=\left.\frac{\partial^{i+j+k} P / T^4}{\partial\hat{\mu}_{\mathrm{B}} ^i \partial\hat{\mu}_{\mathrm{Q}} ^j \partial\hat{\mu}_{\mathrm{S}} ^k}\right|_{\hat{\mu}_{\mathrm{B}, \mathrm{Q}, \mathrm{S}}=0},
\end{align}
where $\hat{\mu}_X\equiv \mu_X/T$ with $X=$ $\{u, d, s\}$ and \{B, Q, S\}.
In this work, we focus on the computation of quadratic fluctuations and correlations, i.e. for $i+j+k=2$. In lattice calculations, we usually calculate the quark fluctuations and then obtain the conserved charge fluctuations via \autoref{eq:BQS_uds_relation}.
Further details can be found in Refs. ~\cite{DElia:2010abb,HotQCD:2012fhj,Bazavov:2017dus,Petreczky:2012rq}.

\subsection{Hadron resonance gas model}

The HRG model assumes that the constituents of the medium are noninteracting pointlike hadrons and resonance states.
Therefore, the pressure of the system in the HRG model can be expressed as the sum of the partial pressures of the individual hadrons and resonances,
\begin{equation}
    \frac{P}{T^4}=\frac{1}{T^4} \sum_i P_i=\frac{1}{V T^3} \sum_i \ln {\mathcal{Z}}_i(V, T, {\mu}_i, B),
    \label{eq:pressure}
\end{equation}
where
\begin{equation}
    \ln{\mathcal{Z}}_i= \pm V g_i \int \frac{\mathrm{~d}^3 \bm{p}}{(2 \pi)^3} \ln \left[1 \pm e^{-\left(E_i-\mu_i\right) / T}\right],
    \label{eq:part_fun}
\end{equation}
represents the grand canonical partition function as a momentum space integral for each resonance $i$. Here $V$ is the volume of the system, $g_i$ is the degeneracy factor, $T$ is the temperature, $E_i$ is the energy of the single particle, and $\mu_i = \mu_{\mathrm{B}}\mathrm{B}_i+ \mu_{\mathrm{Q}}\mathrm{Q}_i+\mu_{\mathrm{S}}\mathrm{S}_i$ is the chemical potential with $\mathrm{B}_i$, $\mathrm{Q}_i$ and $\mathrm{S}_i$ the baryon number, charge and strangeness of the hadron $i$, respectively. The “$+$” in “$\pm$” corresponds to the case for baryons while the “$-$” is for mesons.

\subsubsection{Full space}

In the vanishing magnetic field, the energy of each particle is given by $E_i = \sqrt{|{\bm{p}}|^2+m_i^2}$ with mass $m_i$ and the pressure can be expressed as \cite{HotQCD:2012fhj}
\begin{equation}
    \frac{P_i}{T^4}=\frac{g_i m_i^2}{2(\pi T)^2} \sum_{n=1}^{\infty}( \pm 1)^{n+1} \frac{e^{n \mu_i / T}}{n^2} \mathrm{~K}_2\left(\frac{n m_i}{T}\right),
    \label{eq:HRG_neutral_p}
\end{equation}
where $n$ is the sum index in the Taylor expansion series and $\mathrm{K}_2$ is the second-order modified Bessel function of the second kind.

In the presence of a nonzero magnetic field, we consider all particles to be pointlike, with their masses remaining unaffected by the field. The energy of charged particles is characterized by the Landau level \cite{Endrodi:2013cs},
\begin{equation}
    \varepsilon_i=\sqrt{m_i^{2}+p_{z}^{2}+2\left|q_i\right| B\left(l+1 / 2-s_{z}\right)}\,.
    \label{eq:HRG_energy_level}
\end{equation}
The thermal pressure arising from charged hadrons in nonzero magnetic fields can be expressed as follows \cite{Fukushima:2016vix,Bhattacharyya:2015pra,Ding:2023bft},
\begin{equation}
\begin{aligned}
    \frac{P_{i,charged}}{T^4}=\frac{\left|q_i\right| B}{2 \pi^2 T^3} &\sum_{s_z=-s_i}^{s_i} \sum_{l=0}^{\infty} \varepsilon_T\\
    & \times \sum_{n=1}^{\infty}( \pm 1)^{n+1} \frac{e^{n \mu_i / T}}{n} \mathrm{~K}_1\left(\frac{n \varepsilon_T}{T}\right),
    \label{eq:HRG_charged_p}
\end{aligned}
\end{equation}
where
\begin{equation}
    \varepsilon_T=\sqrt{m_i^{2}+2\left|q_i\right| B\left(l+1 / 2-s_{z}\right)}
    \label{eq:HRG_energy_level_T}
\end{equation}
are the energy levels of charged particles perpendicular to the direction of the magnetic field. Here $q_{i}$ is the electric charge of the hadron $i$,  $s_z$ is the component of the spin in the direction of the magnetic field which is summed over $-s_i$ to $s_i$ for each hadron $i$, $B$ is the magnitude of the magnetic field pointing along $z$ direction, and $l$ denotes the Landau levels.  $\mathrm{K}_1$ is the first-order modified Bessel functions of the second kind. The “$+$” in “$\pm$” corresponds to the case for mesons ($s_{i}$ is integer) while the “$-$” is for baryons ($s_{i}$ is a half integer). 
For neutral particles, we assume that their energy remains unaffected by the magnetic field. Consequently, the pressure they contribute is still given by \autoref{eq:HRG_neutral_p}.

The fluctuations and correlations of conserved charges can be obtained by differentiating the pressure with respect to their corresponding chemical potentials. For charged hadrons, the pressure given in~\autoref{eq:HRG_charged_p} leads to the quadratic fluctuations of and correlations among B, Q, and S arising at zero chemical potential shown as \footnote{Note that the vacuum pressure depends on the magnetic field but is independent of the chemical potential~\cite{Fukushima:2016vix,Ding:2021cwv,Endrodi:2013cs}, thus not contributing to the fluctuations of and correlations among conserved charges.~\autoref{eq:HRG_sus} only needs to consider the contribution of the thermal pressure in ~\autoref{eq:HRG_charged_p}.}\cite{Ding:2021cwv,Ding:2023bft,Fukushima:2016vix}
\begin{equation}
    \begin{aligned}
    &{\chi}_{2}^{X}=\frac{B}{2 \pi^{2} T^3} \sum_{i}\left|q_{i}\right| X_{i}^{2} \sum_{s_{z}=-s_{i}}^{s_{i}} \sum_{l=0}^{\infty} f\left(\varepsilon_{T}\right) \,,\\
    &{\chi}_{11}^{X Y}=\frac{B}{2 \pi^{2} T^3} \sum_{i}\left|q_{i}\right| X_{i} Y_{i} \sum_{s_{z}=-s_{i}}^{s_{i}} \sum_{l=0}^{\infty} f\left(\varepsilon_{T}\right)\,,
    \end{aligned}
    \label{eq:HRG_sus}
\end{equation}
where $f\left(\varepsilon_{T}\right)=\varepsilon_{T} \sum_{n=1}^{\infty}(\pm 1)^{n+1} n \mathrm{~K}_{1}\left(\frac{n \varepsilon_{T}}{T}\right)$ and $X_i,Y_i=$ $\mathrm{B}_i, \mathrm{Q}_i, \mathrm{S}_i$ carried by hadron $i$. 
Note that at a sufficiently strong magnetic field, the lowest energy states of spin-1 mesons and spin-3/2 baryons can become negative, rendering the HRG model inapplicable. This issue will be discussed in detail in Sec. \ref{sec:strong-eB}.
For neutral hadrons, the pressure expressed in~\autoref{eq:HRG_neutral_p} is used to obtain the corresponding fluctuations and correlations, assuming that neutral hadrons are independent of $B$. In our HRG model computations, we have considered all resonances in the QMHRG2020 particle list \cite{ParticleDataGroup:2020ssz}.

\subsubsection{Proxies}
\label{sec:sub_proxies}
In heavy-ion collision experiments, the fluctuations of and correlations among conserved charges are measured through stable, detectable final-state particles. These particles can serve as proxies for conserved charges considering decay chains within the HRG framework \cite{Bellwied:2019pxh}. To align with experimental practice \cite{STAR:2019ans}, we will follow the procedure outlined in \cite{Ding:2023bft}, using net pions ($\tilde{\pi}$), kaons ($\tilde{K}$), and protons ($\tilde{p}$) as proxies for conserved charges:
\begin{align}
    \text{net-B} & \to ~\tilde{p}, \\ 
\text{net-Q} &\to \text{Q}^{\rm PID}\equiv \tilde{\pi}^+ + \tilde{K}^+ +\tilde{p}, \\
\text{net-S} &\to ~ \tilde{K}^+.
\end{align}
Here, we adopt the convention that the net number of particles of a stable species $A$ is defined as $\tilde{A} = A-\bar{A}$, i.e., the difference between the number of particles $A$ and the number of corresponding antiparticles $\bar{A}$.

Within the HRG framework, the proxies corresponding to the conserved charge susceptibilities $\chi^{\rm BQS}_{ijk}$ can be expressed as \cite{Bellwied:2019pxh,Ding:2023bft}
\begin{equation}
    \begin{aligned}
        \sigma_{p,Q^{\rm PID},K}^{i,j,k}=&\sum_R \left(\omega_{R\rightarrow \tilde{p}}\right)^i (\omega_{R\rightarrow \tilde{Q}^{\rm PID}})^j \left(\omega_{R\rightarrow \tilde{K}}\right)^k \\
        &\times I_{\ell}^R\left(T, {\mu}_{\rm B}, {\mu}_{\rm Q}, {\mu}_{\rm S}, e B\right),
    \end{aligned}
\label{eq:proxy}
\end{equation}
where $i+j+k=\ell$ and 
\begin{equation}
    \left.I_\ell^R\left(T, {\mu}_{\rm B}, {\mu}_{\rm Q}, {\mu}_{\rm S}, e B\right)=\frac{\partial^\ell P_R / T^4}{\partial \hat{\mu}_R^\ell}\right|_{\hat{\mu}_{\mathrm{B}, \mathrm{Q}, \mathrm{S}}=0} .
\label{eq:I_Rl}
\end{equation}

The net proxy weights $\omega_{R\rightarrow\tilde{j}} \equiv \omega_{R\rightarrow j} - \omega_{R\rightarrow\bar{j}}$ quantify the average net production of stable hadron $j$ from resonance $R$ decays. These are calculated as
\begin{equation}
    \omega_{R\rightarrow j} = \sum_\alpha \mathcal{N}_{R\rightarrow j}^\alpha n_{j,\alpha}^R,
    \label{eq:Pj_calculation}
\end{equation}
where $\mathcal{N}_{R\rightarrow j}^\alpha$ denotes the branching ratio for decay channel $\alpha$, and $n_{j,\alpha}^R$ represents the multiplicity of particle $j$ produced in that channel. \autoref{eq:Pj_calculation} should account for the stable hadron $j$ generated via the complete decay chain of resonance $R$ \footnote{Within each decay channel $\alpha$, stable particle $j$ can be produced from resonance $R$ through multiple decays, $R\to M_1\to M_2 \cdots\to M_m \to j$, involving multiple intermediate particles $M_i$. Then, the contribution of the entire decay chain is calculated by sequentially multiplying the branching ratios and multiplicities, $\mathcal{N}^{\alpha}_{R\to {M_1}} n_{{M_1,\alpha}}^R \times \left(\prod_{i=1}^{m-1}\mathcal{N}^{\alpha}_{M_i\to M_{i+1}} n_{M_{i+1},\alpha}^{M_i}\right) \times \mathcal{N}^{\alpha}_{{M_m}\to j} n_{j,\alpha}^{M_m}$.}. The electric charge proxy combines contributions,
\begin{equation}
    \omega_{R\rightarrow\tilde{Q}^{\mathrm{PID}}} = \omega_{R\rightarrow\tilde{p}} + \omega_{R\rightarrow\tilde{K}} + \omega_{R\rightarrow\tilde{\pi}}.
    \label{eq:Q_PID}
\end{equation}

Our implementation of proxies includes all decay channels cataloged in the Particle Data Group \cite{ParticleDataGroup:2020ssz} with branching ratios assumed to be independent of magnetic field strength. 
Stable particles ($A$) contribute directly through $\omega_{A \rightarrow A}=1$,
as exemplified by the proton's unit contribution to baryon susceptibility:
$\chi_2^B(T) \ni \omega_{p\rightarrow p} = 1 \quad \text{(proton stability)}.$ 

For instance, we construct the proxies $\sigma_{p}^{2}$, $\sigma_{Q^{\rm PID}}^{2}$, $\sigma_{Q^{\rm PID},p}^{1,1}$, and $\sigma_{Q^{\rm PID},K}^{1,1}$, corresponding respectively to the susceptibilities $\chi_{2}^{\rm B}$, $\chi_{2}^{\rm Q}$, $\chi_{11}^{\rm BQ}$, and $\chi_{11}^{\rm QS}$:
\begin{equation}
\begin{aligned}
    &\sigma_{p}^{2}= \sum_R\left(\omega_{R \rightarrow \tilde{p}}\right)^2 I_2^{R} ,\\
    &\sigma_{Q^{\rm PID}}^{2}= \sum_R\left(\omega_{R \rightarrow \tilde{Q}^{\rm PID}}\right)^2 I_2^{R} ,\\
    &\sigma_{Q^{\rm PID},p}^{1,1}= \sum_R\left(\omega_{R \rightarrow \tilde{p}}\right)\left(\omega_{R \rightarrow \tilde{Q}^{\rm PID}}\right) I_2^{R} \\
    &\sigma_{Q^{\rm PID},K}^{1,1}= \sum_R\left(\omega_{R \rightarrow \tilde{K}}\right)\left(\omega_{R \rightarrow \tilde{Q}^{\rm PID}}\right) I_2^{R} .
\end{aligned}
\label{eq:proxy_pikp}
\end{equation}

\subsubsection{Kinematic cuts}
In high-energy physics experiments, detecting all particles produced in collisions is inherently challenging due to limitations in detector acceptance and resolution. This necessitates the use of kinematic cuts to selectively analyze particles within measurable regions of momentum space. Such cuts become particularly relevant when modeling systems in the presence of external magnetic fields, as implemented in the HRG framework. In this subsection, we outline a method to incorporate these experimental constraints into the HRG model when magnetic fields are present.

Experiments typically define the beam axis as the longitudinal ($z$) axis, with the transverse plane perpendicular to it. Kinematic cuts are then applied using two key observables: the transverse momentum $p_T$ and pseudorapidity $\eta$, defined as 
\begin{equation}
\eta=\left[ \ln  \left(|\boldsymbol{p}|+p_z\right) /\left(|\boldsymbol{p}|-p_z\right)\right] / 2\,.
\end{equation}
Pseudorapidity approximates the spatial rapidity of particles and serves as a proxy for the polar angle relative to the beamline. Different collaborations, such as STAR and ALICE, adopt distinct $p_T$ and $\eta$ thresholds depending on their detector geometries and physics objectives \cite{STAR:2019ans,Saha:2024gdb}. These experiment-specific selection criteria must be explicitly incorporated into theoretical frameworks like HRG to enable meaningful comparison with measurements.

In the standard HRG model without magnetic fields, kinematic cuts are applied by restricting momentum integrals to particles with transverse momentum above a minimum threshold and pseudorapidity within a specified range, effectively limiting the phase space through constrained momentum space integration \cite{Karsch:2015zna},
\begin{equation}
    \int d^3 \boldsymbol{p}=2 \pi \int_{\eta_{\min }}^{\eta_{\max }} d \eta \int_{p_{T_{\min }}}^{p_{T_{\max }}} d p_T p_T|\boldsymbol{p}|.
    \label{eq:pt_cut_d3p_eb0}
\end{equation}
Substituting the above expression into \autoref{eq:pressure}, the pressure of particles in the vanishing magnetic field can be expressed as,
\begin{align}
    \frac{P_{i}^{cuts}}{T^4}&= \frac{g_i}{(2 \pi)^2 T^3} \int_{\eta_{\min}}^{\eta_{\max}} \mathrm{d} \eta \int_{p_{T_{\min }}}^{p_{T_{\max }}} \mathrm{d} p_T |\boldsymbol{p}| p_T \nonumber \\
    &\qquad\qquad \times \sum_{n=1}^{\infty} \frac{(\pm 1)^{n+1}}{n}e^{-n\left(E_i-\mu_i\right)/T} \,.
    \label{eq:pt_cut_eb0}
\end{align}

However, in the presence of an external magnetic field, incorporating kinematic cuts into this framework is not straightforward and requires additional considerations. For collision-generated magnetic fields oriented perpendicular to the reaction plane (along the $y$ axis in the standard coordinate system), charged particle momenta become quantized in the transverse ($x$-$z$) plane via Landau levels:
\begin{equation}
    p_{xz}\left(l, s_y\right) = \sqrt{2 \left|q\right| B\left(l+1 / 2-s_y\right)}\,,
\end{equation}
and the energy levels of particles in the presence of a magnetic field are expressed as,
\begin{equation}
    E_{l}\left(p_y, l, s_y\right)=\sqrt{m^2+p_y^2+2 \left|q\right| B\left(l+1 / 2-s_y\right)}\,.
\end{equation}
Here $s_y$ denotes the spin projection along the direction of the magnetic field. The momentum space integration transforms accordingly:
\begin{equation}
    \begin{aligned}
        \int \mathrm{d}^3 \boldsymbol{p}
        &= \int \mathrm{d}^2 p_{xz} \mathrm{d} p_y
        = \int p_{xz} \mathrm{d} p_{xz} \mathrm{d} p_y \mathrm{d} \phi_p \\
        &= \int \sqrt{2 \left|q_i\right| B\left(l+1 / 2-s_y\right)} \\
        &\qquad \qquad \times ~\mathrm{d} \sqrt{2 \left|q_i\right| B\left(l+1 / 2-s_y\right)} \\
        &\qquad \qquad \times ~\mathrm{d} p_y \mathrm{d} \phi_p\\
        &= \int \left|q_i\right| B \mathrm{d} l \mathrm{d} p_y \mathrm{d} \phi_p\,,
    \end{aligned}
    \label{eq:pt_cut_d3p_nzeB}
\end{equation}
where $\phi_p$ is the azimuthal angle in the $x$-$z$ plane. 
The transverse momentum and pseudorapidity then become functions of Landau-level parameters: 
\begin{equation}
    \begin{gathered}
        p_T = \sqrt{(p_{xz} \sin \phi_p)^2 + p_y^2 }\,,\\ 
        \eta = \text{arcsinh}(p_{xz} \cos \phi_p / p_T)\,.
    \end{gathered}
\end{equation}

Thus, the pressure of charged particles within the framework of HRG in the external magnetic fields can be expressed as,
\begin{equation}
    \begin{aligned}
    \frac{P^{cuts}_{i,charged}}{T^4}=&\frac{\left|q_i\right| B}{(2 \pi)^3 T^3} \sum_{s_y
    =-s_i}^{s_i} \sum_{l=0}^{\infty} \int \mathrm{d} p_y \int \mathrm{d} \phi_p \\
        & \qquad \qquad \qquad \times ~ \Theta(p_{T_{\min}}, p_{T_{\max}}, \eta_{\min}, \eta_{\max}) \\
        & \qquad \qquad \qquad \times ~ \sum_{n=1}^{\infty} \frac{(\pm 1)^{n+1}}{n}e^{-n\left(E_l-\mu_i\right)/T} ,
    \label{eq:charged_kin_cut}
    \end{aligned}
\end{equation}
where the $\Theta(p_{T_{\min}}, p_{T_{\max}}, \eta_{\min}, \eta_{\max})$  function represents the kinematic selection criteria and is formally defined as a product of Heaviside step functions enforcing the experimental cuts:
\begin{equation}
 \begin{aligned}
\Theta(p_{T_{\min}}, p_{T_{\max}}, \eta_{\min}, \eta_{\max}) \equiv &~\Theta(p_T - p_{T_{\min}}) \times \Theta(p_{T_{\max}} - p_T) \\ 
& \times \Theta(\eta - \eta_{\min}) \times \Theta(\eta_{\max} - \eta).
    \end{aligned}
    \label{eq:kcuts_theta}
\end{equation}
\autoref{eq:charged_kin_cut} reduces to the standard magnetic HRG expression~\autoref{eq:HRG_charged_p} when integration limits encompass the full phase space. Therefore, with the help of \autoref{eq:charged_kin_cut} criteria, the numerical evaluation of contribution of charged resonance $i$ within a specific range of $p_T$ and $\eta$ is now feasible. 

In previous subsections, we demonstrated that the HRG susceptibilities \autoref{eq:HRG_sus} and the corresponding proxies \autoref{eq:proxy} account for all individual resonance contributions. However, in experimental analyses, kinematics cuts are typically applied to the stable $\pi$, $K$, and $p$ hadrons \cite{Saha:2024gdb,ALICE:2025mkk, STAR:2019ans}. Therefore, to incorporate these experimental kinematic cuts into the proxies (which consider all resonance decays), we introduce a weight factor $\omega^{cuts}_i$ that encodes the effects of these cuts on the stable particle $i$, defined as
\begin{equation}
    \omega^{cuts}_i = \frac{I^i_\ell(P^{cuts}_i)}{I^i_\ell(P_i)}=\frac{\left.{\partial^\ell (P^{cuts}_i / T^4)}/{\partial \hat{\mu}_i^\ell}\right|_{\hat{\mu}_{\mathrm{B}, \mathrm{Q}, \mathrm{S}}=0}}{\left.{\partial^\ell (P_i / T^4)}/{\partial \hat{\mu}_i^\ell}\right|_{\hat{\mu}_{\mathrm{B}, \mathrm{Q}, \mathrm{S}}=0}},
    \label{eq:cuts_weight}
\end{equation}
where $P^{cuts}_i$ denotes the pressure arising from the particle $i$ with kinematic cuts and $P_i$ is the corresponding pressure without any kinematic cuts. In this work, we mainly adopt  $P^{cuts}_i$ and $P_i$ for charged particles expressed as \autoref{eq:charged_kin_cut} and \autoref{eq:HRG_charged_p}, respectively.\footnote{While for neutral particles, $P^{cuts}_i$ and $P_i$ are given by \autoref{eq:pt_cut_eb0} and \autoref{eq:HRG_neutral_p}, respectively. These expressions also apply to charged particles in the absence of a magnetic field.} And the phase space integral ${I^i_\ell}$ of the leading-order $\ell=2$, with and without kinematic cuts for $\pi$, $K$, and $p$ hadrons is illustrated in \autoref{fig:sup_piKp_kcuts} in Appendix \ref{app:proxy_cuts}.

Consequently, for the proxies of conserved charges, the effects of kinematic cuts---encoded by $\omega^{cuts}_i$---can be translated into the net proxy weights corresponding to the decay of a resonance $R$ to net $\Tilde{\pi}$, $\Tilde{K}$ and $\Tilde{p}$:
\begin{equation}
    \begin{aligned}
        \sigma_{p,Q^{\rm PID},K}^{i,j,k;cuts}=&\sum_R \left(\omega_{R\rightarrow \tilde{p}}~\omega^{cuts}_p\right)^i  \\&\times(\omega_{R\rightarrow\tilde{p}}~\omega^{cuts}_p+ \omega_{R\rightarrow\tilde{K}}~\omega^{cuts}_K+ \omega_{R\rightarrow\tilde{\pi}}~\omega^{cuts}_\pi)^j 
        \\&\times
        \left(\omega_{R\rightarrow \tilde{K}}~\omega^{cuts}_K\right)^k 
        \\&\times 
        I_{\ell}^R\left(T, {\mu}_{\rm B}, {\mu}_{\rm Q}, {\mu}_{\rm S}, e B\right).
    \end{aligned}
\label{eq:proxy_kin_cuts}
\end{equation}
Here, $\sigma_{p,Q^{\rm PID},K}^{i,j,k;cuts}$ represents the proxies for conserved charges that incorporate the experimental kinematic cuts.

In this work, we primarily adopt kinematic cuts corresponding to those used by the ALICE and STAR detectors. Specifically, ALICE~\cite{Saha:2024gdb,ALICE:2025mkk} and STAR~\cite{STAR:2019ans} kinematic cuts are \footnote{ALICE set2 was introduced in a recent experimental analysis~\cite{ALICE:2025mkk}, which appeared on the same day as the first version of our manuscript. This kinematic cut is included in the current version of our analysis.}:
\begin{align}
{\rm ALICE~set}1:~& |\eta| < 0.8,~
\begin{cases}
0.2 < p_T < 2~\text{GeV}/c \quad \text{for } \pi, K \\
0.4 < p_T < 2~\text{GeV}/c \quad \text{for } p
\end{cases}\notag \\
{\rm ALICE~set}2:~& |\eta| < 0.8,~0.4 < p_T < 1.6~\text{GeV}/c \quad \text{for } \pi, K, p \notag \\
{\rm STAR}:~& |\eta| < 0.5,~0.4 < p_T < 1.6~\text{GeV}/c \quad \text{for } \pi, K, p .
\label{eq:kcuts_ALICE_STAR}
\end{align}

\subsection{Ideal gas limit}
\label{subsec:igl}

In the high-temperature limit,  particles can be approximated as a noninteracting ideal gas. Following Ref. \cite{Ding:2021cwv}, the QCD pressure with three flavor massless quarks in the presence of an external magnetic field can be expressed as,
\begin{equation}
    \frac{P}{T^4}=\frac{8 \pi^2}{45}+\sum_{f=u, d, s} \frac{3\left|q_f\right| B}{\pi^2 T^2}\left[\frac{\pi^2}{12}+\frac{\hat{\mu}_f^2}{4}+P_f(B)\right],
    \label{eq:igl_p}
\end{equation}
where
\begin{equation}
   \begin{aligned}
    P_f(B) = \frac{\sqrt{8\left|q_f\right| B}}{T} \sum_{l=1}^{\infty} \sqrt{l} \sum_{k=1}^{\infty} &\frac{(-1)^{k+1}}{k} \cosh \left(k \hat{\mu}_f\right)   \\
    &\times\mathrm{K}_1\left(\frac{k \sqrt{2\left|q_f\right| B l}}{T}\right)\,,
    \end{aligned}
\end{equation}
where $q_f$ denotes the electric charge carried by quark flavor $f$. 
Taking the partial derivative of \autoref{eq:igl_p} with respect to quark chemical potentials, one obtains the fluctuations of and correlations for each quark flavor~\cite{Ding:2021cwv},
\begin{equation}
    \frac{\chi_2^u}{e B}=\frac{4}{\pi^2 T^2}\left(\frac{1}{4}+\hat{b} \sum_{l=1}^{\infty} \sqrt{2 l} \sum_{k=1}^{\infty}(-1)^{k+1} k \mathrm{~K}_1(k \hat{b} \sqrt{2 l})\right)\,,
    \label{eq:igl_2u}
\end{equation}
\begin{equation}
    \frac{\chi_2^{d, s}}{e B}=\frac{2}{\pi^2 T^2}\left(\frac{1}{4}+\hat{b} \sum_{l=1}^{\infty} \sqrt{l} \sum_{k=1}^{\infty}(-1)^{k+1} k \mathrm{~K}_1(k \hat{b} \sqrt{l})\right)\,,
    \label{eq:igl_2d_2s}
\end{equation}
\begin{equation}
    \chi_{11}^{ud}=\chi_{11}^{us}=\chi_{11}^{ds}=0\,.
\end{equation}

In the limit $\sqrt{eB}/T \rightarrow \infty$, \autoref{eq:igl_2u} and \autoref{eq:igl_2d_2s} retain contributions only from their first terms. The corresponding values of the second-order fluctuations of and correlations among conserved charges are listed in \autoref{tab:free_limit}.

\begin{table}[!ht]
    \centering
    \begin{tabular}{cccc}
        \hline \hline
        ~ &      B       &       Q      &       S      \\
        \hline 
        B & $2/(9\pi^2)$ & $1/(9\pi^2)$ &$-1/(6\pi^2)$ \\
        Q & $1/(9\pi^2)$ & $5/(9\pi^2)$ & $1/(6\pi^2)$ \\
        S &$-1/(6\pi^2)$ & $1/(6\pi^2)$ & $1/(2\pi^2)$ \\
        \hline \hline	
    \end{tabular}
    \caption{The values of second order fluctuations and correlations of conserved charge normalized by the magnetic field strength and multiplied by the square of the temperature, $\chi_{11}^{XY}T^2/eB$ and $\chi_{2}^{X}T^2/eB = \chi_{11}^{XX}T^2/eB$ with $X,Y=\rm B,Q,S$ in the free limit with $\sqrt{eB}/T\rightarrow \infty$\cite{Ding:2021cwv}.}
    \label{tab:free_limit}
\end{table}


\section{Lattice setup}
\label{sec:lattice_setup}
The QCD partition function can be written as
\begin{equation}
    \mathcal{Z} = \int \mathcal{D} U \prod_{f = u,d,s}^{}[\det M_f(U,q_f B,m_f)]^{{1}/{4}} e^{-S_g},
\end{equation}
where $M_f(U,q_f B,m_f)$ denotes the fermion matrix for flavor $f$, and $S_g$ represents the gauge action. 
Our simulations employ 2+1 flavor highly improved staggered quarks (HISQ) \cite{Follana:2006rc} and a tree-level improved Symanzik gauge action, consistent with the framework used by the HotQCD Collaboration \cite{Bazavov:2019www,Bollweg:2021vqf,Bollweg:2022rps,Bollweg:2024epj}. 

A uniform magnetic field oriented along the $z$ direction is implemented through $U(1)$ link variables $u_{\mu}(n)$ in the Landau gauge \cite{Bali:2011qj, AlHashimi:2008hr},
\begin{equation}
    \begin{aligned}
    & u_x\left(n_x, n_y, n_z, n_\tau\right)= \begin{cases}\exp \left[-i a^2 q B N_x n_y\right] & \left(n_x=N_x-1\right) \\
    1 & \text { (otherwise) }\end{cases} \\
    & u_y\left(n_x, n_y, n_z, n_\tau\right)=\exp \left[i a^2 q B n_x\right], \\
    & u_z\left(n_x, n_y, n_z, n_\tau\right)=u_t\left(n_x, n_y, n_z, n_\tau\right)=1 ,
    \end{aligned}
\end{equation}
where the lattice dimensions are $(N_x,~ N_y,~ N_z,~ N_\tau )$ with $n_{\mu}=0,...~,N_\mu - 1(\mu = x,~y,~z,~\tau)$.

In this setup, the strength of the magnetic field, $eB$, is quantized and can be expressed as follows \cite{Bali:2011qj,Ding:2023bft},
\begin{equation}
    eB = \frac{6 \pi N_{b}}{N_{x} N_{y}} a^{-2}, \quad N_b \in \mathbb{Z}\,,
\end{equation}
where $N_{b}$ counts magnetic flux quanta through the $x$-$y$ plane and $a$ is the lattice spacing. The electric charges of the quarks follow $q_d=q_s=-q_u/2=-e/3$ with $e$ being the elementary electric charge. To maintain flux quantization across all flavors, we chose the greatest common divisor of electric charges, i.e., $\left|q_{d}\right|=\left|q_{s}\right|=e /3$, in our implementation. A periodic boundary condition for U(1) links is imposed for all directions except for the $x$ direction, constraining $N_b$ to the range $0\leq N_b<\frac{N_x N_y}{4}$. Our setup features $N_{\sigma} \equiv N_x=N_y=N_z$ with aspect ratio $N_{\sigma}/N_{\tau}=4$. Further details on the implementation of magnetic fields with HISQ fermions can be found in Ref.~\cite{Ding:2020hxw}. In this work, simulations are mainly performed on $32^3 \times 8$ and $48^3 \times 12$ lattices in nonzero magnetic fields. Additionally, we have conducted simulations on a $64^3\times 12$ lattice at fixed $N_b$ and temperature to assess the uncertainty arising from the continuum extrapolation and estimates. To mitigate discretization effects associated with magnetic fields, it is required that $a^2 qB \ll 1$, which translates to $N_b / N_\sigma^2 \ll 1$. In the literature, $N_b / N_\sigma^2 < 5\%$ is commonly adopted \cite{DElia:2021tfb,Endrodi:2019zrl}. In our simulations, the largest magnetic flux $N_b = 32$ on a $32^3 \times 8$ lattice corresponds to a maximum $N_b / N_\sigma^2$ of about $3\%$, which remains within the accepted range, ensuring that discretization effects from the magnetic field are mild. Notably, this ratio becomes even smaller on the $48^3 \times 12$ lattice.

\begin{figure}[!htp]    
    \centering
    \includegraphics[width=0.4\textwidth]{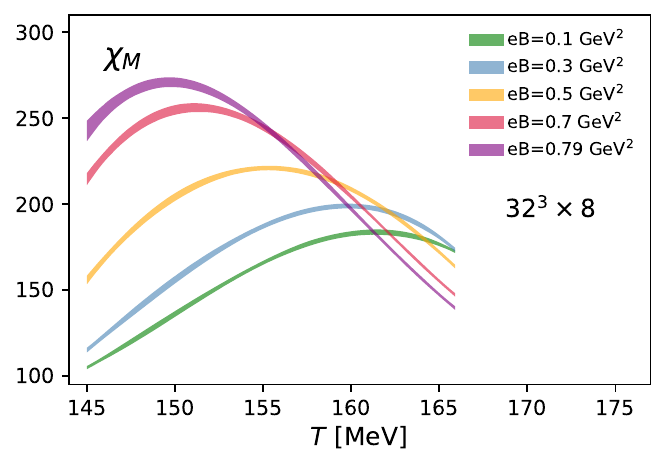}
    \includegraphics[width=0.4\textwidth]{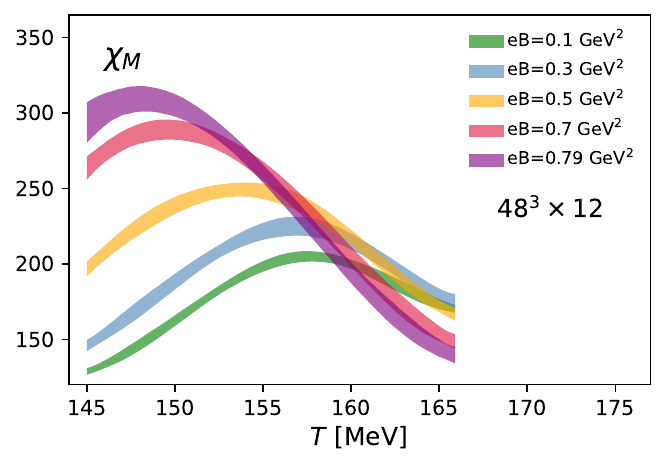}
    \caption{Chiral susceptibility $\chi_{M}(eB)$ as a function of temperature for several fixed magnetic field strengths $eB$, computed on lattices with temporal extents $N_\tau = 8$ (top) $ N_\tau = 12$ (bottom).} 
    \label{fig:chi_M}
\end{figure}

The current simulations improve upon our earlier studies in two key aspects: first, the quark masses are fixed to their physical values with degenerate light quarks, yielding a pion mass of 135 MeV at $eB=0$, in contrast to the $M_\pi\simeq 220$ MeV configuration used in Ref.~\cite{Ding:2021cwv}. Second, we extend the magnetic flux range to include $N_b$ values of $1, 2, 3, 4, 6, 12, 16, 24$ and 32, significantly surpassing the smaller $N_b$ values explored in Ref.~\cite{Ding:2023bft}, which only considered $N_b=1 - 6$. The details on the simulation parameters can be found in Appendix~\ref{app:stat}. The lattice QCD results in the absence of a magnetic field are taken from Ref.~\cite{Bollweg:2021vqf}.


\section{The transition line on $T$-$eB$ plane}
\label{sec:Tpc-eB}
To determine the crossover transition temperature $T_{pc}$ at nonzero magnetic fields, we analyze the peak location of total chiral susceptibility $\chi_M(eB)$, which is defined as the quark mass derivative of chiral condensate. We use the following subtracted chiral condensate $M$ and its corresponding susceptibility $\chi_M$~\cite{Bazavov:2018mes,Ding:2023bft},

\begin{equation}
    M =\frac{1}{f_K^4}\left[m_s\left(\langle\bar{\psi} \psi\rangle_u+\langle\bar{\psi} \psi\rangle_d\right)-\left(m_u+m_d\right)\langle\bar{\psi} \psi\rangle_s \right] ,
\end{equation}
\begin{equation}
\begin{aligned}
    \chi_M &=\left.m_s\left(\partial_{m_u}+\partial_{m_d}\right) M\right|_{m_u=m_d=m_l} \\
    &=\frac{1}{f_K^4} \left [m_s\left(m_s \chi_l-2\langle\bar{\psi} \psi\rangle_s-4 m_l \chi_{s u}\right) \right ],
\end{aligned}
\end{equation}
where $\langle\bar{\psi} \psi\rangle_f={T}\left({\partial \ln \mathcal{Z}}/{\partial m_f}\right)/{V}$ with $f=u,d,s$ for the up, down and strange quark, respectively. Here, $f_K=155.7(9) / \sqrt{2}~ \mathrm{MeV}$ is the kaon decay constant, $\chi_{f g}=\partial_{m_f}\langle\bar{\psi} \psi\rangle_g$, and $\chi_l=\chi_{u u}+\chi_{u d}+\chi_{d u}+\chi_{d d}$. Note that at nonzero magnetic fields, $\chi_{uu}$ and $\chi_{dd}$ become nondegenerate, and mixed susceptibilities $\chi_{ud}$ ($\chi_{du}$) are explicitly included.

\begin{figure}[!htp]    
    \centering
    \includegraphics[width=0.4\textwidth]{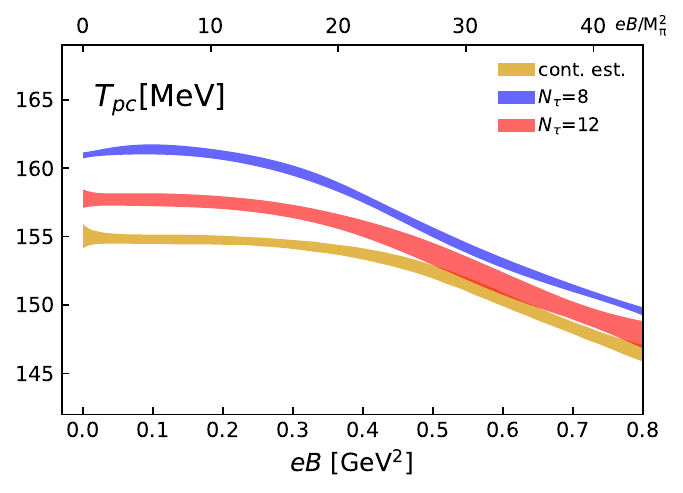}
    \caption{Pseudocritical temperature $T_{pc}(eB)$ of the QCD chiral crossover, determined from the peak locations of the chiral susceptibility as a function of magnetic strength $eB$. Continuum estimates (gold bands) were obtained using the corresponding results for $ N_\tau =8$ (blue bands) and $ N_\tau =12$ (red bands).} 
    \label{fig:Tpc}
\end{figure}

\begin{figure*}[!htbp]    
\centering
\includegraphics[width=0.32\textwidth]{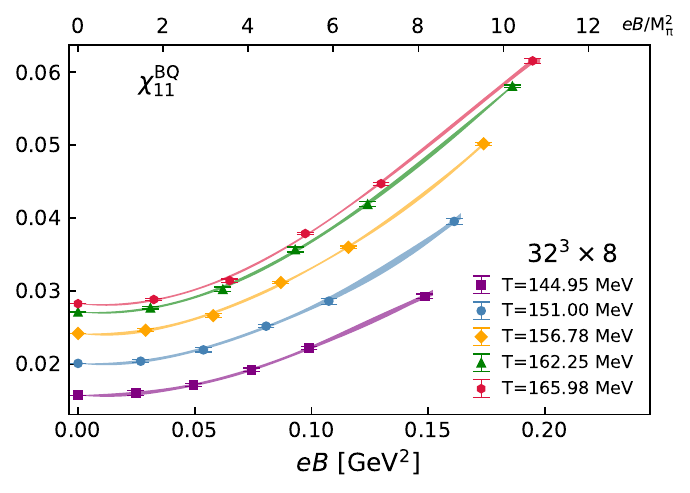}
\includegraphics[width=0.32\textwidth]{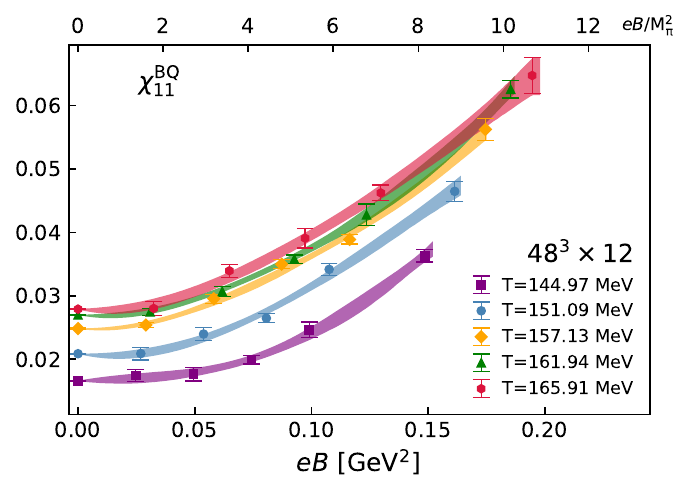}
\includegraphics[width=0.32\textwidth]{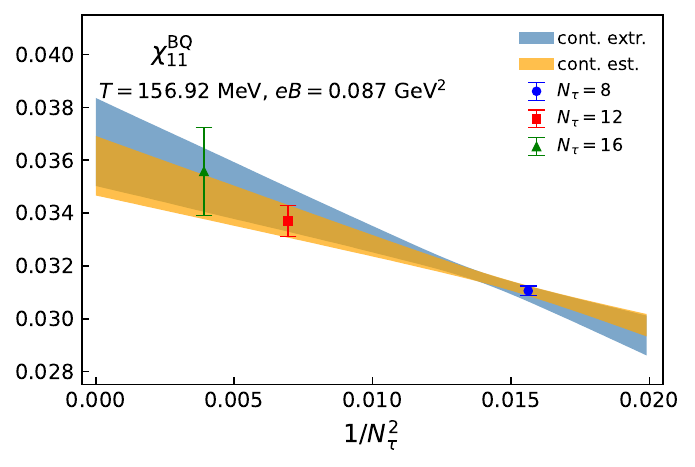}
\caption{Baryon-electric charge correlations $\chi^{\rm BQ}_{11}$ obtained from lattice QCD simulations on lattices with temporal extents $N_{\tau} = 8$ (left) and $N_{\tau} = 12$ (middle), spanning $T$-$eB$ parameter space in the relatively weak-$eB$ regime. The colored bands represent two-dimensional spline interpolations of lattice data points (filled symbols) at different fixed temperatures. The right panel shows continuum estimates of \chiBQ obtained from linear extrapolations in $1/N_{\tau}^2$ using lattice data at $N_\tau=8$ and $12$, as well as continuum-extrapolated results incorporating additional $N_\tau=16$ data, evaluated at a fixed value of $eB$.
}
\label{fig:ldata_chiBQ11_est}
\end{figure*}

Since the magnetic field does not affect the UV-divergent part \cite{Bali:2011qj}, in practice, we determine the peak location of the following $\chi_M(eB)$ as $T_{pc}(eB)$:
\begin{equation}
    \begin{aligned}
        \chi_M(eB) &=&\frac{m_s}{f_K^4}\Big[m_s \chi_l(eB)-2\langle\bar{\psi} \psi\rangle_s(eB=0) \nonumber \\ 
       &&\quad -4 m_l \chi_{s u}(eB=0)\Big] .
    \end{aligned}
\end{equation}

We utilize five temperature points spanning $T=145~{\rm MeV}$ and $T=166~{\rm MeV}$ and perform spline fits to locate the susceptibility peak. \autoref{fig:chi_M} shows the magnetic field dependence of chiral susceptibility on $32^3\times8$ and $48^3\times12$ lattices. It can be observed that the peak location of the chiral susceptibility shifts to lower temperatures with increasing $eB$ across both lattice spacings.

In Fig.~\ref{fig:Tpc}, the continuum estimated $T_{pc}(eB)$, obtained through an extrapolation ansatz that we will detail in the following section, reveals two distinct regimes: for $eB \lesssim 0.3~\mathrm{GeV}^2$, $T_{pc}$ exhibits only a mild dependence on the magnetic field, while for $eB > 0.3~\mathrm{GeV}^2$, $T_{pc}$ decreases monotonically with increasing $eB$.  This behavior aligns with earlier studies \cite{Bali:2011qj,DElia:2021yvk}, which determined the transition temperature using different observables, such as the renormalized chiral condensate etc.

In the following analysis, we systematically investigate the fluctuations of conserved charges in two distinct magnetic field regimes. First, we examine relatively weak magnetic fields with $N_b = 1, 2, 3, 4, 6$, corresponding to $eB \lesssim 0.15~\mathrm{GeV}^2$. Subsequently, we explore the behavior at strong magnetic fields with $N_b = 12, 16, 24, 32$, extending the analysis up to $eB \simeq 0.8~\mathrm{GeV}^2$.


\section{Results in relatively weak magnetic fields}
\label{sec:weak-eB}

We begin by presenting lattice data for baryon-electric charge correlations, $\chi^{\rm BQ}_{11} \equiv \chi^{\rm BQ}_{11}(T,eB)$ and their continuum estimates in \autoref{fig:ldata_chiBQ11_est}. The left and middle panels show results for $N_\tau=8$ and $N_\tau=12$ lattices, respectively, across our $T$-$eB$ parameter space. The discrete magnetic field values arise from flux quantization in finite volumes, necessitating spline interpolation.\footnote{Further studies utilizing polynomial fitting and parameterization will be presented in future research on the equation of state in strong magnetic fields \cite{Eos}.} Similar to the approach mentioned in Ref. \cite{Bali:2011qj}, we perform two-dimensional B-spline fits to the lattice data. The average values and error bands are obtained using the Gaussian bootstrap method, with the median and $68\%$ percentiles of the distribution, respectively. The interpolated data bands for $N_{\tau} = 8$ (left) and $N_{\tau} = 12$ (middle) in \autoref{fig:ldata_chiBQ11_est} are depicted in distinct colors, corresponding to results at fixed temperatures across varying external magnetic field strengths.

The right panel in \autoref{fig:ldata_chiBQ11_est} illustrates continuum estimates, determined through linear extrapolations in $1/N_\tau^2$ using the $ N_\tau = 8 $ and $ N_\tau = 12 $ interpolated data. For validation, we include a supplementary $ N_\tau = 16 $ data point and perform an independent continuum extrapolation incorporating $ N_\tau = 8 $, $ N_\tau = 12 $, and $ N_\tau = 16 $ results under the same $1/N_\tau^2 $ ansatz. The continuum estimated results and continuum extrapolated results from both extrapolation procedures exhibit consistency within uncertainties. Given this agreement, we adopt the continuum estimates derived from $ N_\tau = 8 $ and 12 lattices for subsequent analyses, as they provide a statistically robust foundation while minimizing computational overhead.

\begin{figure*}[t]    
\centering
\includegraphics[width=0.32\textwidth]{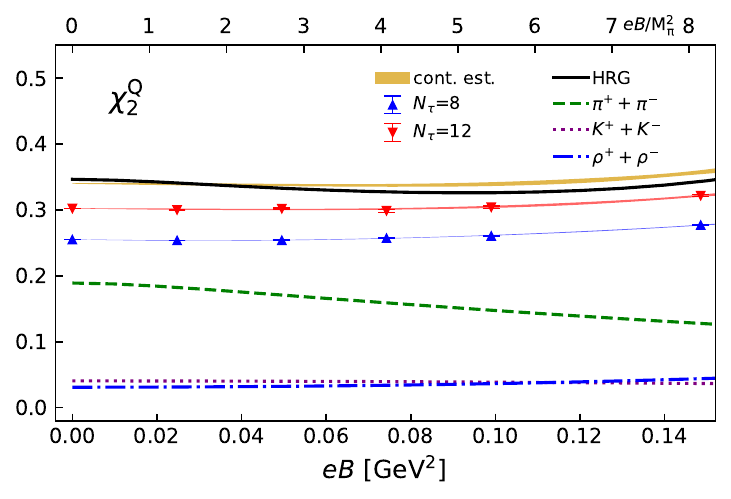}
\includegraphics[width=0.32\textwidth]{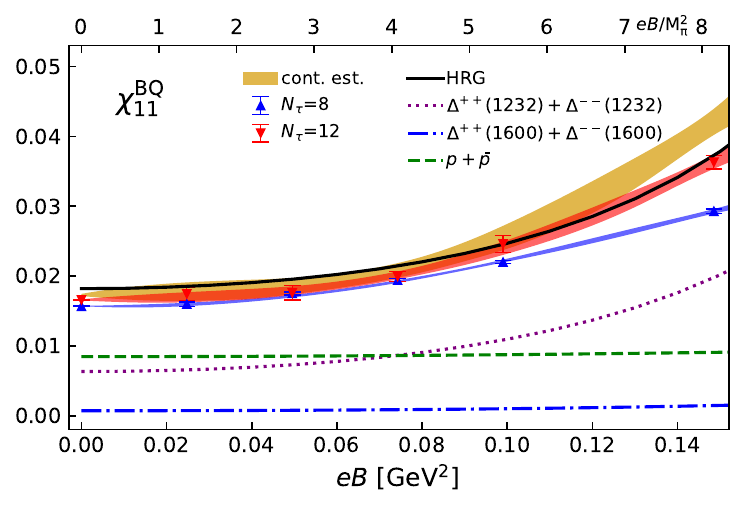}
\includegraphics[width=0.32\textwidth]{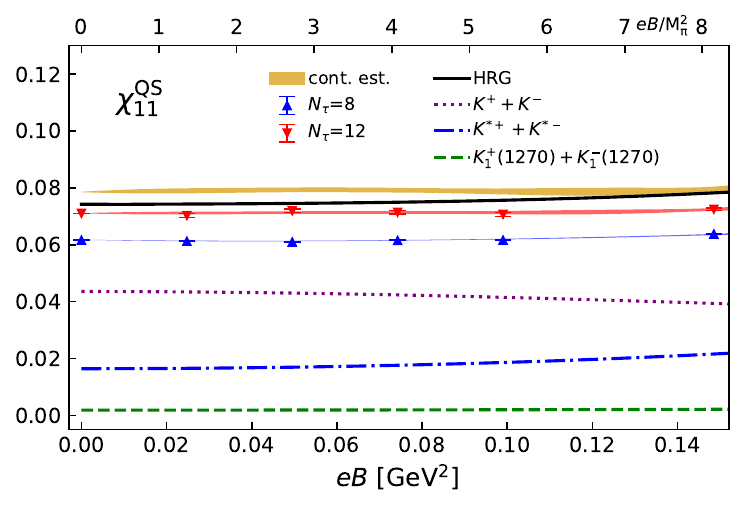}
\caption{Electric charge fluctuations $\chi^{\rm Q}_{2}$ (left) and their correlations with baryon number $\chi^{\rm BQ}_{11}$ (middle) and strangeness $\chi^{\rm QS}_{11}$ (right) at $T=145~{\rm MeV}$ as functions of magnetic strength $eB$ in the relatively weak-$eB$ regime. Continuum estimates (gold bands) were obtained using the interpolated $ N_\tau =8$ (blue) and $ N_\tau =12$ (red) bands and corresponding simulation data points (filled symbols). Black solid and colored broken lines represent the total and dominant-resonance contributions, respectively, within the HRG model. }
\label{fig:charge-suscp-T145}
\end{figure*}

\begin{figure*}[!htbp]
\centering
\includegraphics[width=0.32\textwidth]{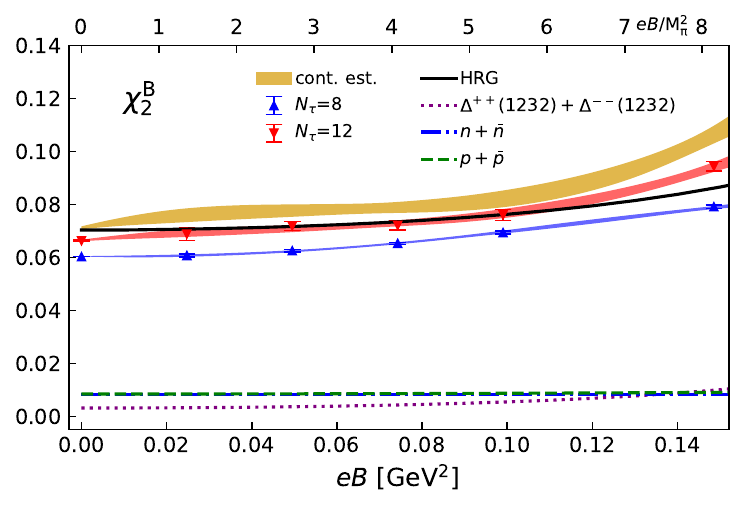}
\includegraphics[width=0.32\textwidth]{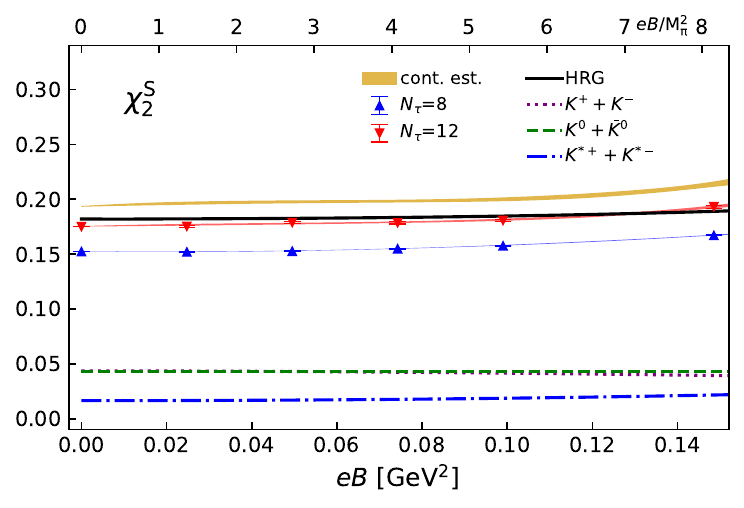}
\includegraphics[width=0.32\textwidth]{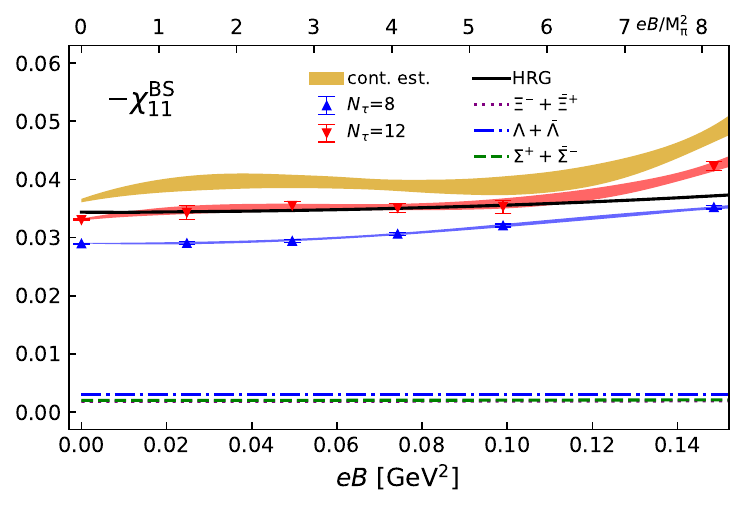}
\caption{Same as Fig. \ref{fig:charge-suscp-T145}, but for the fluctuations of baryon number $\chi^{\rm B}_{2}$ (left) and strangeness $\chi^{\rm S}_{2}$ (middle), and corresponding correlations $-\chi^{\rm BS}_{11}$ (right) among them.}
\label{fig:neutral-suscp-T145}
\end{figure*}

The existence of sustained primordial magnetic fields and their potential imprints in heavy-ion collisions have driven extensive investigations into electric charge fluctuations, particularly to probe topological phenomena such as the chiral magnetic effect. While the interplay between electric charge and magnetic fields might intuitively suggest a pronounced enhancement of electric charge fluctuations at all baryon chemical potentials, this expectation is not borne out in practice. One key advantage of studying electric charge fluctuations ($\chi^{\rm Q}_{2}$) and their correlations with baryon number ($\chi^{\rm BQ}_{11}$) and strangeness ($\chi^{\rm QS}_{11}$) lies in their theoretical tractability: these quantities derive exclusively from charged hadrons, whose energy spectra in magnetic fields are comparatively well understood \cite{Bali:2017ian,Ding:2020hxw}. In contrast, contributions from neutral hadrons---whose magnetic field-dependent energy modifications remain less well established and are often approximated as constant---can introduce uncertainties in other susceptibilities, such as baryon number ($\chi^{\rm B}_{2}$), strangeness ($\chi^{\rm S}_{2}$), or baryon-strangeness correlations ($\chi^{\rm BS}_{11}$).

\autoref{fig:charge-suscp-T145} presents lattice QCD results and HRG model predictions for the susceptibilities $\chi^{\rm Q}_{2}$, $\chi^{\rm BQ}_{11}$, and $\chi^{\rm QS}_{11}$ at $T = 145\,\mathrm{MeV}$, a temperature below the pseudocritical temperature $T_{ pc}(eB)\approx 155\,\mathrm{MeV}$, which remains nearly constant for magnetic fields $eB \lesssim 0.3\,\mathrm{GeV}^2$, as shown in~\autoref{fig:Tpc}. We focus initially on these susceptibilities since their exclusive dependence on charged hadrons enables direct and theoretically consistent comparisons between lattice QCD and HRG calculations. Susceptibilities involving neutral hadrons will be discussed subsequently.
Interpolated results for $N_\tau = 8$ (blue bands) and $N_\tau = 12$ (red bands) lattices are displayed alongside continuum estimates (gold bands). Contrary to naive expectations, the continuum estimate for $\chi^{\rm Q}_{2}$ remains largely unaffected by $eB$, while $\chi^{\rm BQ}_{11}$ exhibits striking sensitivity. Although $\chi^{\rm Q}_{2}$ dominates in magnitude at low $T$ among \chiBQ~and \chiQS ---reflecting predominant mesonic contributions---it exhibits little sensitivity to $eB$, likely due to cancellations between baryonic and mesonic sectors. Meanwhile, $\chi^{\rm QS}_{11}$ shows negligible sensitivity to $eB$, consistent with the large strange quark mass suppressing magnetic field-driven effects.

At low temperatures, such as $T=145~{\rm MeV}$, the HRG model is expected to effectively describe the QCD-driven behavior of fluctuations of conserved charges at vanishing magnetic fields. However, as discussed in the previous section, the presence of magnetic fields complicates this picture due to modifications in the hadron spectrum. For charged hadrons, these modifications are quantified by the Landau-level-dependent energy spectrum (see \autoref{eq:HRG_energy_level}). The magnetized HRG model (black solid lines in \autoref{fig:charge-suscp-T145}) provides a reasonable description of $\chi^{\rm Q}_{2}$ and $\chi^{\rm BQ}_{11}$ in the weak-field regime ($eB \lesssim 4\,M_{\pi}^2$), but begins to undershoot lattice QCD results at stronger fields. For $\chi^{\rm QS}_{11}$, however, a persistent discrepancy exists: the HRG model systematically underestimates lattice data across the weak-$eB$ regime, with only coincidental agreement emerging at higher $eB$.

To elucidate the differences between HRG predictions and lattice results, we analyze the dominant contributions of individual charged hadrons.  The $eB$-dependence of these contributions arises from Landau quantization, where the energy levels of charged hadrons shift with $eB$ depending on their spin projection $s_z$ (aligned or antialigned with the field). For $\chi^{\rm Q}_{2}$, charged pions dominate the contributions, analogous to the zero-field case. However, their contribution decreases by approximately $30\%$ at $eB \simeq 8\,M_{\pi}^2$, likely due to the magnetic field-induced increase in charged pion mass \cite{Ding:2020hxw}. Contributions from kaons and $\rho$ mesons, though enhanced with $eB$, remain much smaller in magnitude compared to those from pions.

\begin{figure*}[!htbp]    
\centering
\includegraphics[width=0.32\textwidth]{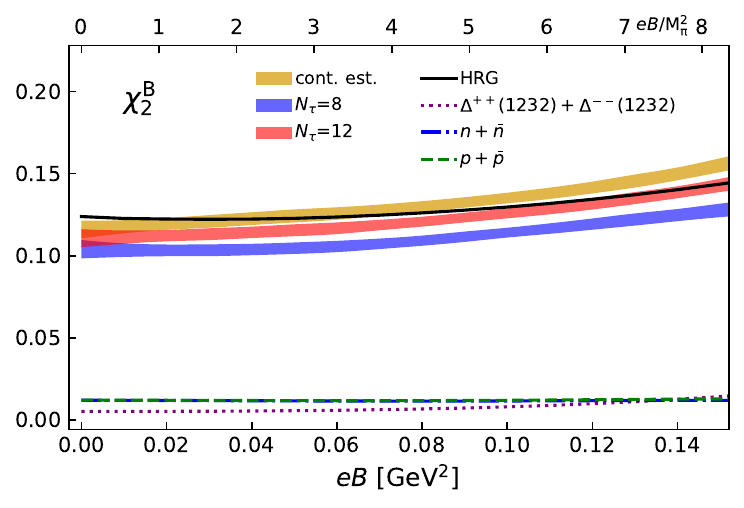}
\includegraphics[width=0.32\textwidth]{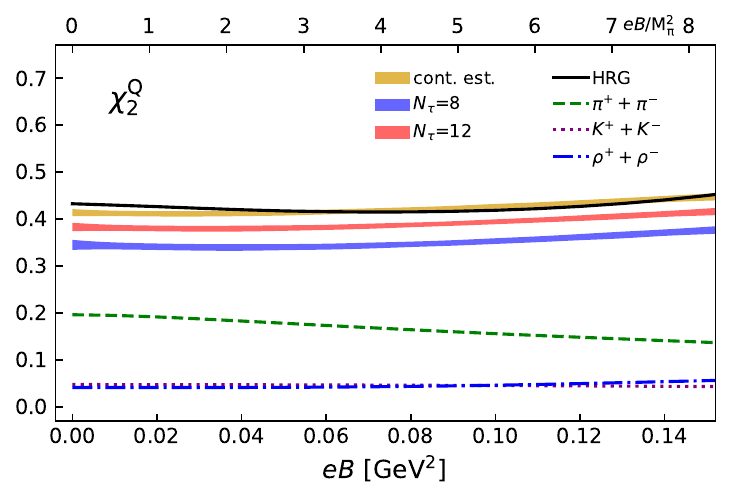}
\includegraphics[width=0.32\textwidth]{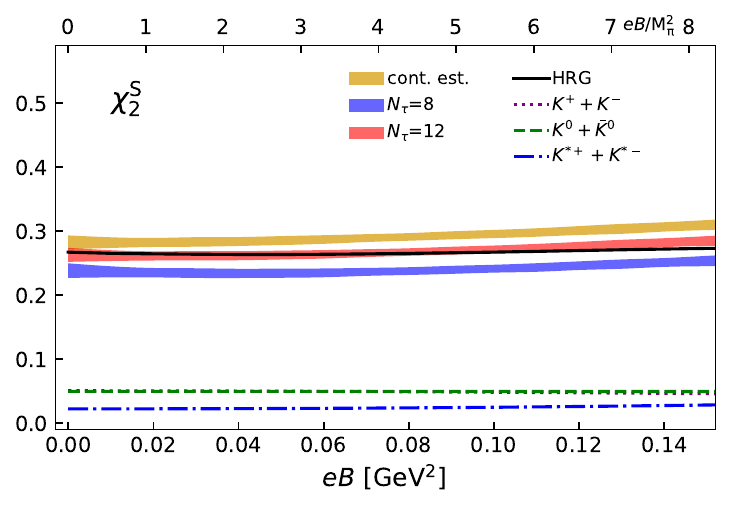}
\includegraphics[width=0.32\textwidth]{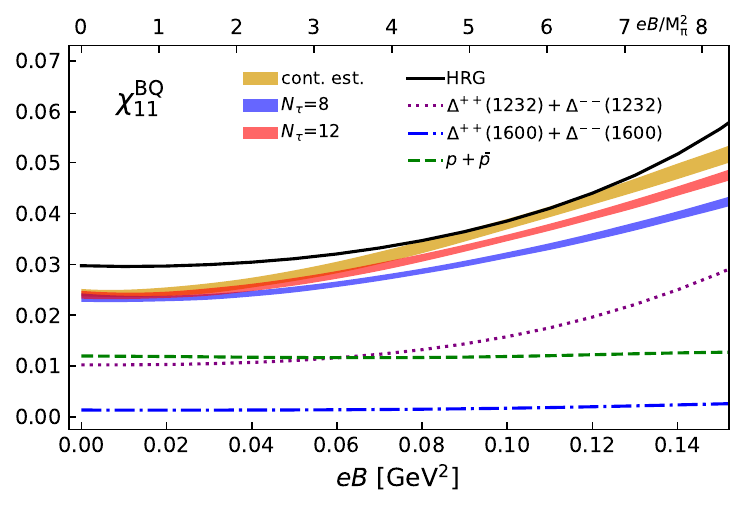}
\includegraphics[width=0.32\textwidth]{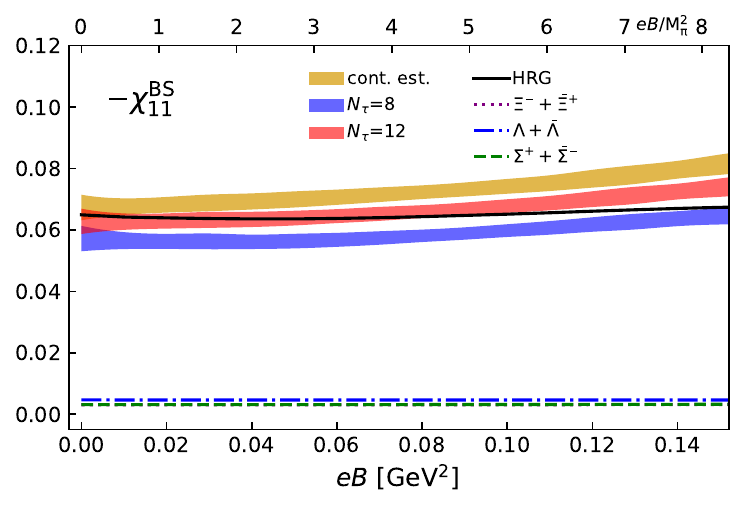}
\includegraphics[width=0.32\textwidth]{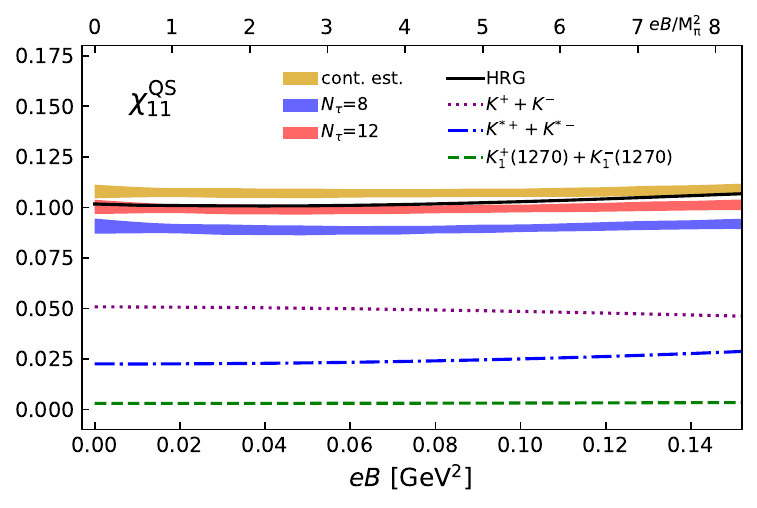}
\caption{Fluctuations of and correlations among conserved charges, $\chi^{\rm B}_{2}$, $\chi^{\rm Q}_{2}$, $\chi^{\rm S}_{2}$, $\chi^{\rm BQ}_{11}$, $-\chi^{\rm BS}_{11}$ and $\chi^{\rm QS}_{11}$ (top left to bottom right), along the transition line $T_{pc}(eB)$ in the relatively weak-$eB$ regime. Continuum estimates (gold bands) were obtained using the interpolated $ N_\tau =8$ (blue bands) and $ N_\tau =12$ (red bands) results along $T_{pc}(eB)$. Black solid and colored broken lines represent the total and dominant-resonance contributions, respectively, within the HRG model.}
\label{fig:allsus_Tpc}
\end{figure*}

For $\chi^{\rm BQ}_{11}$, protons dominate at $eB = 0$, but the doubly charged $\Delta(1232)$ baryons become the primary contributors near $eB \simeq 4\,M_{\pi}^2$. Notably, the proton contribution remains nearly constant with $eB$, while the $\Delta(1232)$ contribution grows due to its spin-dependent mass reduction for specific $s_z$ states---overcoming the suppression from states with increased mass. Heavier doubly charged baryons, such as $\Delta(1600)$, remain exponentially suppressed due to their large masses.

For $\chi^{\rm QS}_{11}$, charged pseudoscalar $K$ mesons dominate the contributions throughout the $eB$-regime, with their contributions declining gradually as $eB$ increases, while the vector $K^*$ mesons exhibit a modest rise. The overall insensitivity of $\chi^{\rm QS}_{11}$ to $eB$ reflects the weaker magnetic response of kaons compared to pions~\cite{Ding:2020hxw}, stemming from the interplay between the magnetic field and the spin states of the hadrons.

Next, we turn to fluctuations involving neutral hadrons. In \autoref{fig:neutral-suscp-T145}, we present lattice QCD computation and the HRG model results for $\chi^{\rm B}_{2}$ (left), $\chi^{\rm S}_{2}$ (middle) and $\chi^{\rm BS}_{11}$ (right) at $T=145 ~{\rm MeV}$.  While $\chi^{\rm B}_{2}$ exhibits a moderate enhancement of $\sim45\%$ at $eB \simeq 8\,M_{\pi}^2$, this increase is notably smaller than the pronounced sensitivity observed for $\chi^{\rm BQ}_{11}$. The HRG model systematically underestimates $\chi^{\rm B}_{2}$ across the entire $eB$ range. This discrepancy may arise from unaccounted magnetic field effects on neutrons: studies show that neutral hadrons like pions and kaons exhibit mass reductions in magnetic fields \cite{Bali:2017ian,Ding:2020hxw} at low temperature, and similar effects might apply to neutrons. A decrease in neutron mass would amplify their contributions to $\chi^{\rm B}_{2}$, potentially bridging the gap between HRG predictions and lattice data. However, in our current HRG framework, neutron masses are held constant with $eB$, resulting in almost equal contributions from neutrons and protons. The proton’s contribution to $\chi^{\rm B}_{2}$ mirrors its behavior in $\chi^{\rm BQ}_{11}$ (see middle panel of~\autoref{fig:charge-suscp-T145}), remaining flat with $eB$. Only at the upper end of the $eB$ range ($eB \simeq 8\,M_{\pi}^2$) do doubly charged $\Delta(1232)$ baryons overtake nucleons as the dominant contributors, driven by spin-aligned states with reduced masses.  

In contrast, $\chi^{\rm S}_{2}$ and $\chi^{\rm BS}_{11}$ show negligible sensitivity to $eB$, though their mechanisms may differ. The HRG model underestimates both quantities, similar to the discrepancies observed in $\chi^{\rm QS}_{11}$. This underestimation of  $\chi^{\rm S}_{2}$ and $\chi^{\rm BS}_{11}$ from HRG may stem partly from the unaccounted magnetic field effects on neutral hadrons. For $\chi^{\rm S}_{2}$, the dominant contributions arise from the isospin doublets $\{K^+, K^0\}$ and $\{K^-, \bar{K}^0\}$, with charged kaons showing stable contributions despite $eB$, while neutral kaons remain static. For $\chi^{\rm BS}_{11}$, no single hyperon species dominates, reflecting the weaker interplay between strangeness and baryon number in magnetic fields.

As the temperature approaches the pseudocritical value $T_{pc}(eB)$, the conserved charge susceptibilities exhibit systematic increases across all six second-order measures, as shown in ~\autoref{fig:allsus_Tpc}. This monotonic increase in susceptibilities is likely due to the fact that $ T_{pc}(eB)$ remains nearly constant with $eB$, as discussed earlier. If $T_{pc}(eB) $ were to decrease significantly with $eB$, the susceptibilities might exhibit nonmonotonic behavior in temperature.  The $eB$-dependence of these susceptibilities mirrors trends observed at $T = 145~{\rm MeV}$: $\chi^{\rm Q}_{2}$ remains largely insensitive to $eB$, while $\chi^{\rm B}_{2}$ shows a reduced enhancement compared to its low-temperature behavior. Similarly, $\chi^{\rm QS}_{11}$, $\chi^{\rm S}_{2}$, and $\chi^{\rm BS}_{11}$ exhibit minimal sensitivity to $eB$, consistent with their behavior at lower temperatures. Notably, $\chi^{\rm BQ}_{11}$, the most $eB$-sensitive observable, retains its pronounced enhancement, further solidifying its utility as a magnetic field probe.

The hadron resonance gas model, while formally inapplicable at $T_{pc}(eB)$, accidentally aligns with lattice QCD results for certain susceptibilities in specific regimes---particularly for $eB \lesssim 4\,M_\pi^2$ as seen in~\autoref{fig:allsus_Tpc}. This unexpected agreement highlights the need for refined studies in weaker magnetic fields where $T_{pc}(eB) \approx T_{pc}(0)$. A decomposition of dominant hadronic contributions reveals parallels to the low-temperature case, with one key distinction: the transition from proton-dominated to $\Delta(1232)$-dominated contributions in $\chi^{\rm BQ}_{11}$ shifts to lower $eB$ ($\sim 3\,M_{\pi}^2$) at $T_{pc}(eB)$, compared to $eB \simeq 4\,M_{\pi}^2$ at $T = 145~{\rm MeV}$. This suggests enhanced sensitivity of higher-spin resonances to thermal and magnetic effects near the transition.  Additionally, $\chi^{\rm BS}_{11}$ shows an increased contribution from $\Lambda$ hyperons at $T_{pc}(eB)$, indicating subtle thermal modifications to strangeness-baryon correlations. These findings highlight the complex interplay between magnetic field strength and hadronic degrees of freedom across the QCD phase diagram.

\begin{figure}[!htbp]    
\centering
\includegraphics[width=0.45\textwidth]
{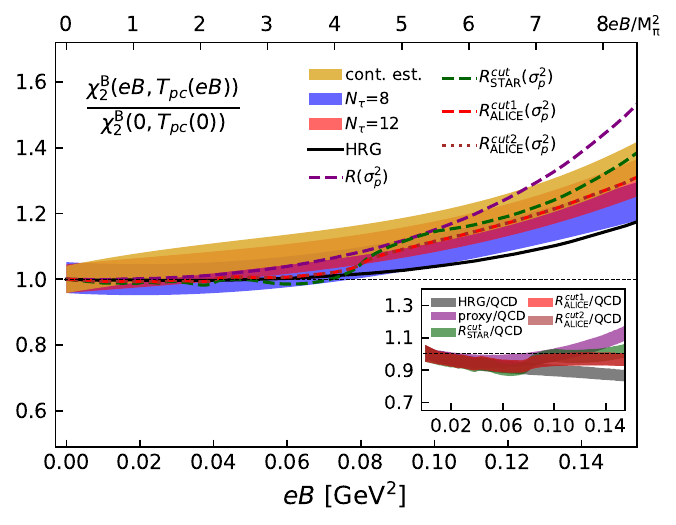}
\includegraphics[width=0.45\textwidth]
{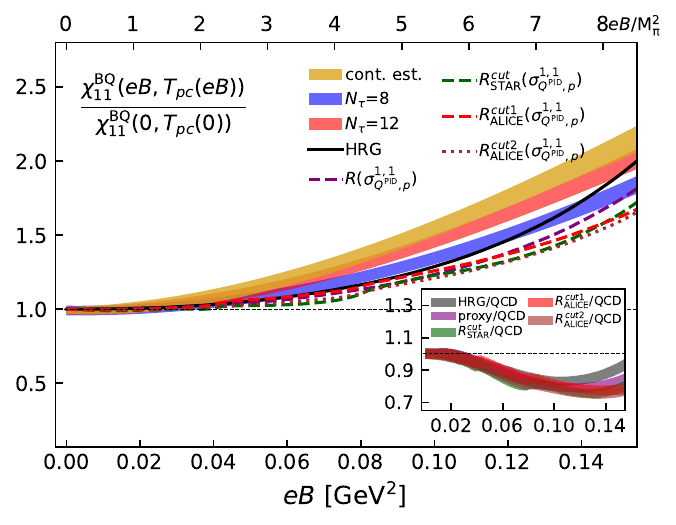}
\caption{The ratio observables $R(\chi^{\rm B}_{2})$ (top) and $R(\chi^{\rm BQ}_{11})$ (bottom) along the transition line $T_{pc}(eB)$. Continuum estimates (gold bands) were obtained using the interpolated $ N_\tau =8$ (blue bands) and $ N_\tau =12$ (red bands) results along $T_{pc}(eB)$. The solid line denotes results calculated from the HRG model, while the dashed lines correspond to experimental proxies: the purple, green, and red dashed lines represent proxies integrated over the full phase space, and those with kinematic cuts applied for the STAR and ALICE detectors, respectively. The insets depict the ratio of the HRG model and experimental proxies to the corresponding lattice QCD continuum estimates.}
\label{fig:rcp_Tpc}
\end{figure}

The conserved charge susceptibilities discussed above are directly related to their experimental counterparts \cite{Luo:2017faz,Pandav:2022xxx}, expressed as cumulants of the net-charge distributions $N_X$,
\begin{align}
\chi^{\rm X}_{2}(T,eB) &= \frac{1}{VT^3}\langle \left(N_X - \langle N_X \rangle \right)^2 \rangle\,, \\
\chi^{\rm XY}_{11}(T,eB) &= \frac{1}{VT^3}\langle \left(N_X - \langle N_X \rangle \right)\left(N_Y - \langle N_Y \rangle \right) \rangle\,.
\end{align}

However, these cumulants inherently include system volume effects, which are not directly measurable. To address this limitation, both theoretical and experimental studies often advocate the use of ratio observables to suppress volume-dependent contributions \cite{Skokov:2012ds,Gavai:2010zn,STAR:2019ans}. In our analysis, it is natural to examine the behavior of observables at finite $eB$ relative to those at vanishing magnetic fields. This approach is analogous to studying peripheral-to-central collision conditions, as encapsulated in $R_{cp}$-like observables. Specifically, we define the ratio observable as 
\begin{equation}
 R(\mathcal{O}) \equiv \mathcal{O}(eB,T)/  \mathcal{O}(eB=0,T).
\end{equation}
This formulation allows us to focus on the effects of the magnetic field while minimizing the influence of system volume uncertainties.

\autoref{fig:rcp_Tpc} presents the ratios $R(\mathcal{O})$  for \chitB~ and \chiBQ~ along the transition line $T = T_{pc}(eB)$. As shown in~\autoref{fig:allsus_Tpc} these two observables exhibit the strongest sensitivity to $eB$ at $T_{pc}(eB)$, motivating our focus on them here. In addition to the suppression of volume effects, we observe a notable reduction in lattice artifacts, i.e. the $N_{\tau} = 8,~12$ interpolated data bands align more closely with the continuum estimates.

It can be seen from \autoref{fig:rcp_Tpc} that both $R(\chi^{\rm B}_{2})$ and $R(\chi^{\rm BQ}_{11})$ successfully capture the magnetic effects along the transition line $T_{pc}(eB)$, reaching values of about 1.25 and 2.1, respectively, for $eB \simeq 8\,{M_\pi}^2$. These substantial enhancements, particularly the pronounced response of \chiBQ~compared to \chitB,  reiterate the magnetometer proposition of $\chi^{\rm BQ}_{11}$ \cite{Ding:2023bft}.

The HRG model, represented by black solid lines in \autoref{fig:rcp_Tpc}, lacks rigorous validity at the transition temperature. HRG predictions for individual susceptibilities diverge from lattice data as seen in \autoref{fig:allsus_Tpc}: \chiBQ~overshoots lattice values, while \chitB ~aligns reasonably (albeit possibly coincidentally) along the transition line. However, when normalized via the ratio $R(\mathcal{O})$---which suppresses absolute magnitudes while amplifying trends in $eB$-dependence---the model exhibits partial agreement with lattice results.  Notably, the $eB$-dependence of the normalized ratios $R(\mathcal{O})$, where amplitude effects cancel, is qualitatively reproduced by HRG for $eB \lesssim 0.04 \, \text{GeV}^2$ ($\sim 2.5\,M_\pi^2$) as seen in \autoref{fig:rcp_Tpc}. Beyond this regime, HRG systematically underestimates the magnitude of $R$(\chitB), yielding an HRG-to-lattice ratio of $\sim 0.85$ at $eB \simeq 8\,M_\pi^2$. For $R$(\chiBQ), the HRG-to-lattice ratio decreases to a minimum of 0.83 at $eB \simeq 0.1 \, \text{GeV}^2$ before recovering toward unity. While HRG fails to quantitatively reproduce the absolute values of individual susceptibilities or their ratios, it accounts for $\sim 80\%$ of the $eB$-dependent trends in the normalized observables. This suggests that HRG retains limited predictive power for the relative behavior of fluctuations under weak magnetic fields, even at $T_{pc}$, despite its inapplicability in absolute terms.

While the HRG model captures approximate trends in normalized observables under weak magnetic fields, its connection to experimental observables requires further consideration. As discussed in Section~\ref{sec:sub_proxies}, particles carrying baryon number and electric charge in heavy-ion collisions are not all directly detectable. We therefore construct proxies for $R(\chi^{\rm B}_{2})$ and $R(\chi^{\rm BQ}_{11})$, defined as $R(\sigma^{2}_{p})$ (proton variance) and $R(\sigma^{1,1}_{Q^{\rm PID},p})$ (charge-proton correlations with identified protons/pions/kaons), respectively (see~\autoref{eq:proxy_pikp}).  For $eB \lesssim 2.5\,M_\pi^2$ ($\sim 0.04$ GeV$^2$), both proxies of $R(\sigma^{2}_{p})$ and $R(\sigma^{1,1}_{Q^{\rm PID},p})$ remain near unity, consistent with HRG predictions (insets, \autoref{fig:rcp_Tpc}). 
However, as $eB$ grows, $R(\sigma^{2}_{p})$ deviates monotonically from HRG results and reaches to about 1.3 times the HRG results at $eB \simeq 8\,M_\pi^2$.

When compared to QCD results, $R(\sigma^{2}_{p})$ agrees well up to $eB \sim 5\,M_\pi^2$, and start to exceed QCD results at $eB\sim 7M_\pi^2$, reaching $\sim 1.1$ times QCD results at $eB \simeq 8\,M_\pi^2$ (top panel, inset). In contrast, $R(\sigma^{1,1}_{Q^{\rm PID},p})$ aligns more closely to the HRG results, and only starts to deviate from the HRG results at $eB\gtrsim 5M_\pi^2$. The deviation increases as $eB$ grows, and reaches to $10\%$ at the highest field. The origin of this discrepancy lies in the construction of $R(\sigma^{1,1}_{Q^{\rm PID},p})$, which relies solely on correlations among final-state protons, kaons and pions. As demonstrated in~\autoref{fig:charge-suscp-T145} and~\autoref{fig:allsus_Tpc}, the $eB$-dependence of \chiBQ~ is predominantly driven by the doubly charged $\Delta(1232)$ resonance. Since the $\Delta(1232)$ is unstable and decays via $\Delta^{++} \to p + \pi^+$ with a near-100\% branching ratio, its decay products (proton and pion) do not fully capture the $eB$-dependence of $\Delta(1232)$'s contribution to \chiBQ. Despite this limitation, $R(\sigma^{1,1}_{Q^{\rm PID},p})$ effectively captures a substantial portion of the $eB$-dependence. Compared to QCD results, it deviates by $\sim 22\%$ at $eB \simeq 0.1$ GeV$^2$ and $\sim 10\%$ at $eB \simeq 8\,M_\pi^2$ (bottom panel, inset). Thus, the proxy $R(\sigma^{1,1}_{Q^{\rm PID},p})$ still captures about 80\% of the $eB$-dependence observed in lattice QCD, only missing roughly 20\% of the full $eB$ sensitivity.

\begin{figure*}[!htbp]    
\centering
\includegraphics[width=0.4\textwidth]{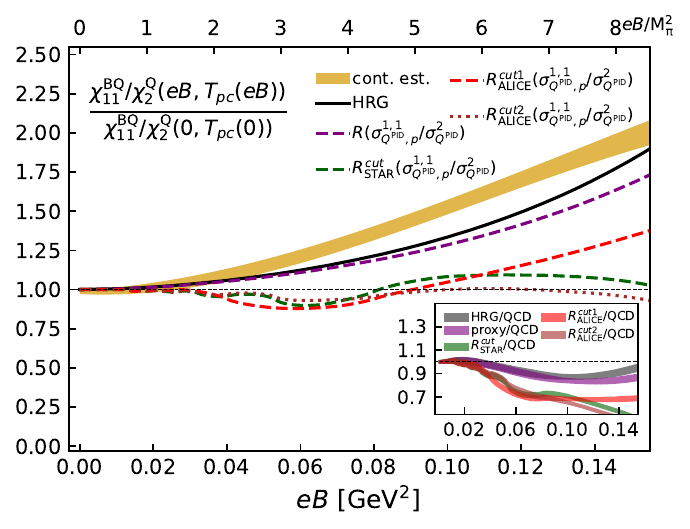}
\includegraphics[width=0.4\textwidth]{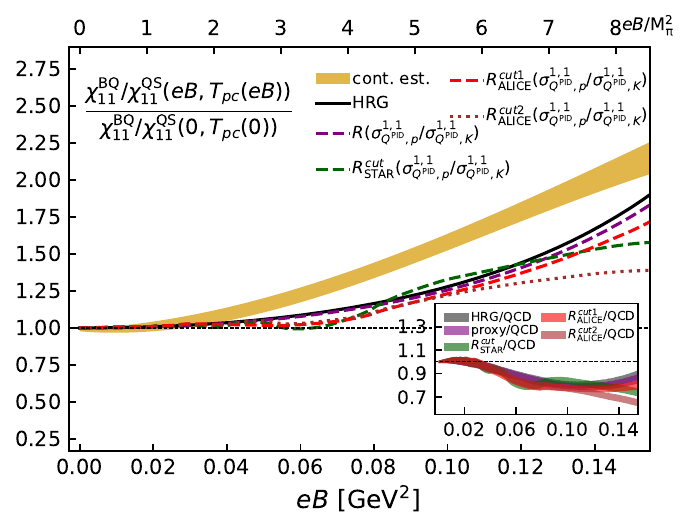}
\includegraphics[width=0.4\textwidth]{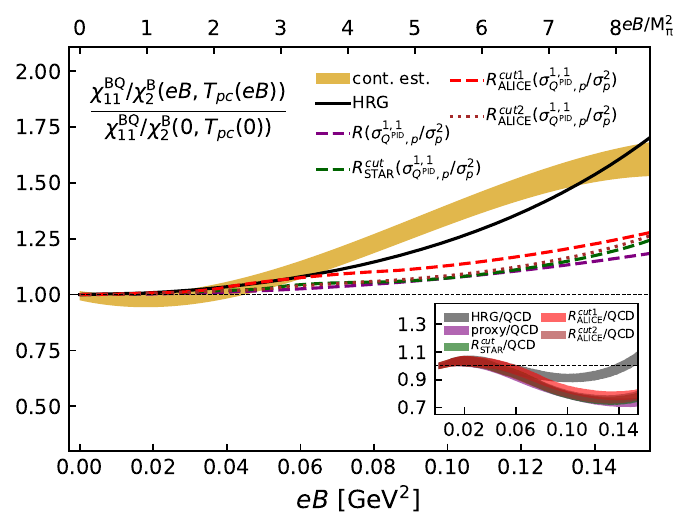}
\includegraphics[width=0.4\textwidth]{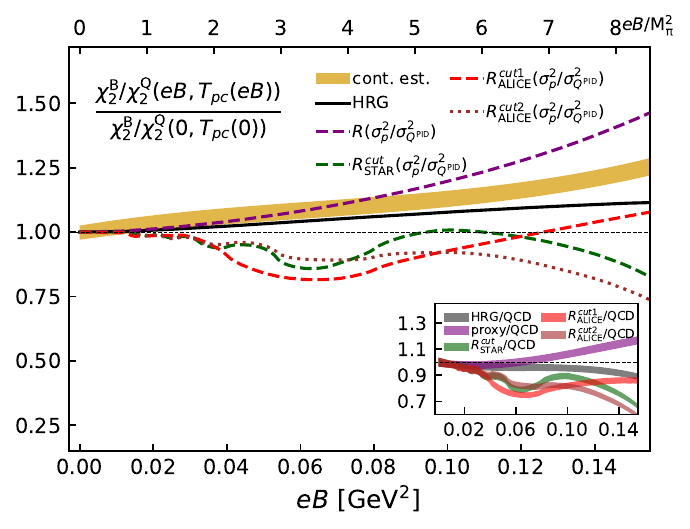}
\caption{Double ratios $R(\chi^{\rm BQ}_{11}/\chi^{\rm Q}_{2})$, $R(\chi^{\rm BQ}_{11}/\chi^{\rm QS}_{11})$, $R(\chi^{\rm BQ}_{11}/\chi^{\rm B}_{2})$ and $R(\chi^{\rm B}_{2}/\chi^{\rm Q}_{2})$ (from top to bottom) along $T=T_{pc}(eB)$. The gold band represents the continuum estimates obtained from lattice computation. The solid line denotes results calculated from the HRG model, while the dashed lines correspond to experimental proxies: the purple, green, and red dashed lines represent proxies integrated over the full phase space, and those with kinematic cuts applied for the STAR and ALICE detectors, respectively. The insets depict the ratio of the HRG model and experimental proxies to the corresponding lattice QCD continuum estimates.}
\label{fig:double_rcp_Tpc}
\end{figure*}

To align with experimental practicality, we further construct detector-specific proxies,  $R^{cut}_{\rm STAR}$, $ R^{cut1}_{\rm ALICE}$ and $ R^{cut2}_{\rm ALICE}$, incorporating kinematic cuts reflective of the STAR and ALICE detectors, respectively. These results, alongside HRG and lattice comparisons, are also illustrated in~\autoref{fig:rcp_Tpc}. Generally, these proxies with kinematic cuts yield smaller values compared to the full phase space results. The difference between the kinematically constrained results and those obtained in full phase space is notably smaller for \chiBQ~than for \chitB, and the results with kinematic cuts applied to the ALICE and STAR detectors demonstrate similar sensitivity levels.
Specifically, for \chitB~the proxy measure $R(\sigma^{2}_{p})$ somewhat exceeds the lattice QCD results. The introduction of detector-specific kinematic cuts (ALICE and STAR) reduces this overshoot, aligning the proxies more closely with lattice predictions. Similarly, for $R$(\chiBQ), proxies constrained by realistic detector kinematics still capture approximately 80\% of the lattice-predicted sensitivity, missing about 20\%. These findings indicate that, despite the observed partial loss, the experimental proxies for $R$(\chitB) and $R$(\chiBQ)~maintain sufficient sensitivity to magnetic field effects, supporting their effectiveness for detecting and quantifying relevant signatures in heavy-ion collision experiments.

Building upon the results discussed above, and motivated by both the demonstrated utility of $R_{cp}$-like ratio observables and the need to systematically compare $eB$-driven enhancements across susceptibilities, we introduce double ratio $R_{cp}$-like observables,
\begin{equation}
     R(\mathcal{O}_2/\mathcal{O}_1) \equiv \frac{\mathcal{O}_2(eB,T)/\mathcal{O}_1(eB,T)}{\mathcal{O}_2(eB=0,T)/\mathcal{O}_1(eB=0,T)}.
\end{equation}
Such ratios, widely adopted in heavy-ion experiments to suppress volume effects \cite{Saha:2024gdb,STAR:2019ans}, are typically analyzed as functions of centrality. This is motivated by the fact that magnetic field strength grows from central to peripheral collisions, creating a correlation between centrality and $eB$ in collision systems. To isolate magnetic field effects, we prioritize combinations of $\mathcal{O}_2/\mathcal{O}_1$ that maximize sensitivity to $eB$. As established earlier, \chiBQ~and \chitB~ exhibit the strongest $eB$-dependence along $T_{pc}(eB)$, while other second-order susceptibilities (namely \chitQ, \chitS, \chiQS, \chiBS ) show milder responses.

To amplify magnetic field signatures we thus select $\mathcal{O}_2$ as \chiBQ~or \chitB~and $\mathcal{O}_1$ as one of the less sensitive susceptibilities in the double ratio $R(\mathcal{O}_2/\mathcal{O}_1)$~\footnote{Although \chiQS, \chiBS~ and \chitS~ exhibit mild $eB$-sensitivities, we nevertheless provide $R$(\chiQS/\chitS) and $R$(\chiBS/\chitS) along with corresponding proxies in \autoref{fig:app_double_rcp} in Appendix~\ref{app:proxy_cuts} in comparison with preliminary results from the ALICE Collaboration~\cite{Saha:2024gdb}.}.
For demonstration, we choose $\mathcal{O}_1 =$\chitQ~or \chiQS~ retaining experimental accessibility while isolating $eB$-driven deviations.  We show in the top two panels of \autoref{fig:double_rcp_Tpc} the double ratios of \chiBQ~to \chitQ~and \chiQS~along the transition line. These two double ratios start to deviate from unity at $eB \simeq 0.02$ GeV$^2$, and then increase rapidly as $eB$ grows, reaching values of $\sim 2$ at the largest magnetic fields. In both cases, the HRG results agree with lattice data when the ratios remain near 1 at $eB \lesssim 0.02$ GeV$^2$, but begin to undershoot the lattice data as $eB$ increases. The largest deviation between HRG and lattice results is $\sim 20\%$ at $eB \simeq 0.1$ GeV$^2$, with the discrepancy for \chiBQ/\chiQS~being slightly larger. On the other hand, the proxy obtained in the full phase space generally shows similar sensitivity on $eB$ as the HRG results do. 

When incorporating the kinematic cuts corresponding to the STAR and ALICE detectors, the experimental proxies for $R$(\chiBQ/\chitQ), denoted as $R^{cut}_{\rm STAR}$, $R^{cut1}_{\rm ALICE}$ and $R^{cut2}_{\rm ALICE}$, exhibit significant suppression compared to HRG and lattice QCD results, as seen from the top left plot in~\autoref{fig:double_rcp_Tpc}. Proxies with both kinematic cuts remain about unity at $eB\lesssim$ 0.02 GeV$^2$ and start to form a dip around $eB\simeq 0.06$ GeV$^2$ and subsequently rise above unity for $eB\gtrsim$ 0.09 GeV$^2$ ($\sim 5\,M_\pi^2$). Notably, at stronger magnetic fields, $eB\gtrsim$ 0.09 GeV$^2$, $R^{cut}_{\rm STAR}$ only marginally exceeds unity before returning to unity at our largest simulated $eB$, while $R^{cut2}_{\rm ALICE}$ remains almost consistent with unity. In contrast, $R^{cut1}_{\rm ALICE}$ continues to increase as $eB$ grows and reaches about 1.25 at our largest $eB$. This 25\% enhancement observed in $R^{cut1}_{\rm ALICE}$ aligns closely to the recently reported measurements by the ALICE collaboration using set1 kinematic cuts at the most peripheral collisions \cite{ALICE:2025mkk}.

\begin{figure*}[!htp]    
\centering
\includegraphics[width=0.32\textwidth]{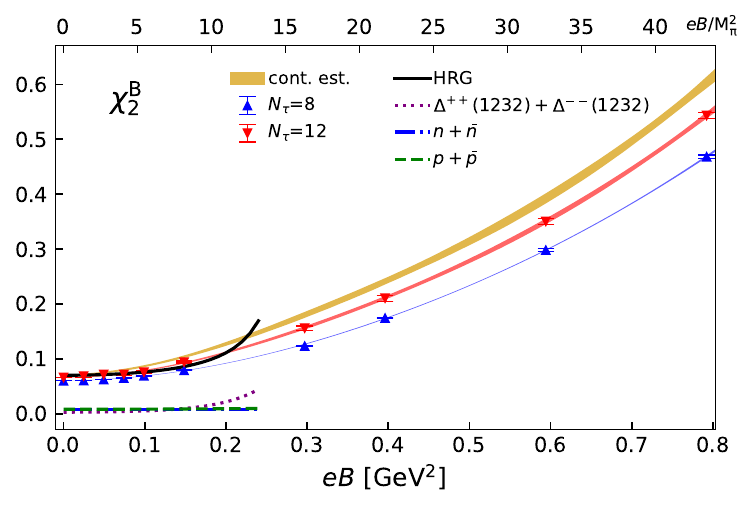}
\includegraphics[width=0.32\textwidth]{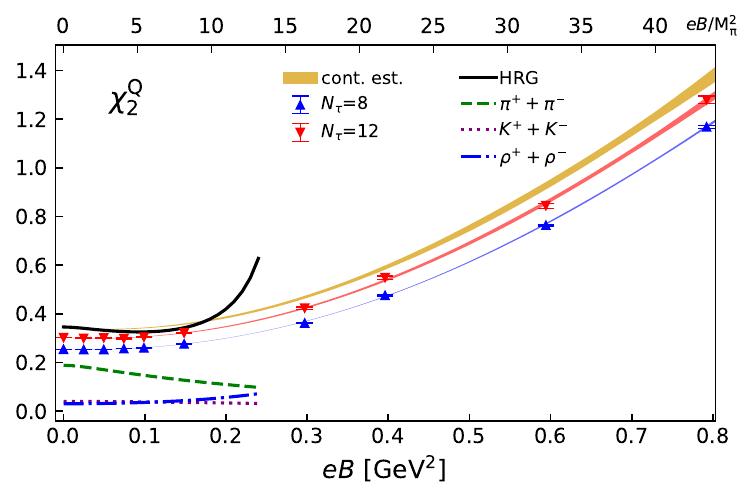}
\includegraphics[width=0.32\textwidth]{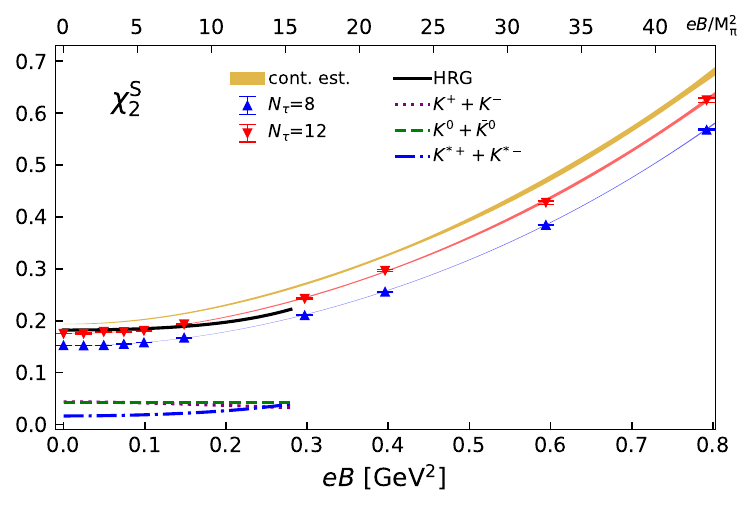}
\includegraphics[width=0.32\textwidth]{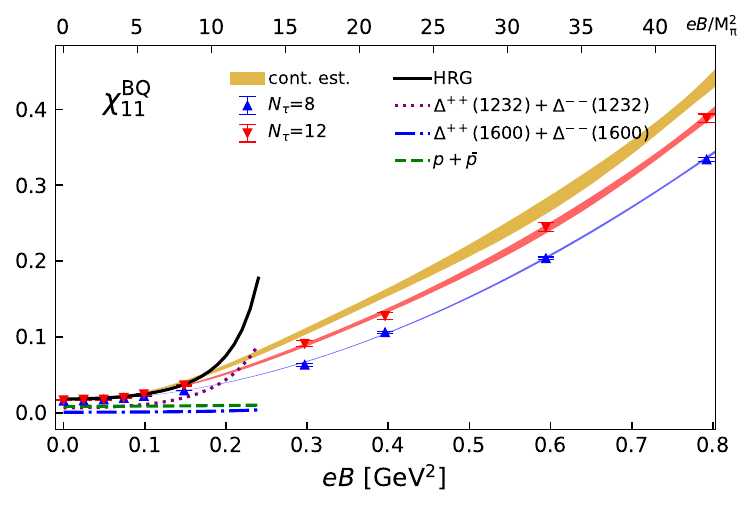}
\includegraphics[width=0.32\textwidth]{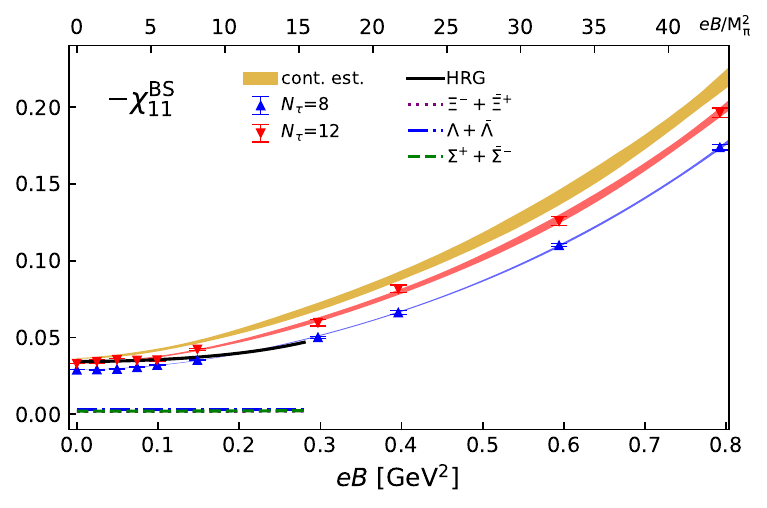}
\includegraphics[width=0.32\textwidth]{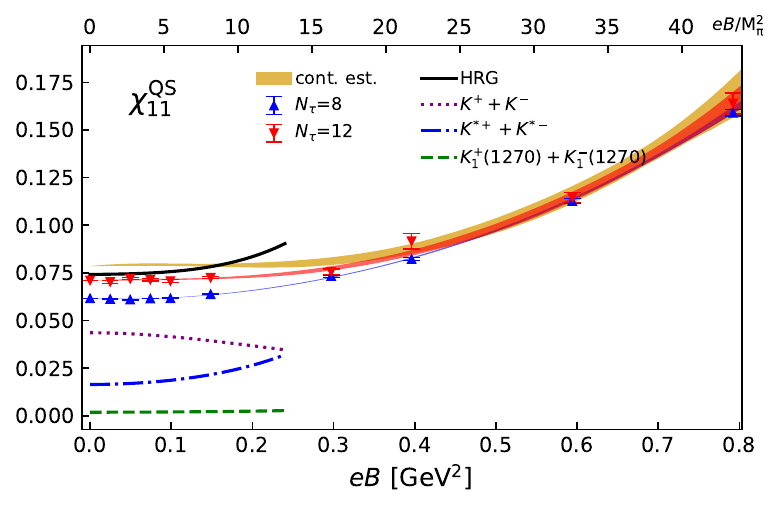}
\caption{Fluctuations of and correlations among conserved charges, $\chi^{\rm B}_{2}$, $\chi^{\rm Q}_{2}$, $\chi^{\rm S}_{2}$, $\chi^{\rm BQ}_{11}$, $-\chi^{\rm BS}_{11}$ and $\chi^{\rm QS}_{11}$ (top left to bottom right),  at $T=145~{\rm MeV}$ versus $eB$ up to the strong-$eB$ regime. Continuum estimates (gold bands) were obtained using the interpolated $ N_\tau =8$ (blue) and $ N_\tau =12$ (red) bands and corresponding simulation data points (filled symbols). Black solid and colored broken lines represent the total and dominant-resonance contributions, respectively, in the HRG model.}
\label{fig:suscp-vseB_strong_T145}
\end{figure*}

As seen in the top right panel of \autoref{fig:double_rcp_Tpc}, $R$(\chiBQ/\chiQS) exhibits a sensitivity to $eB$ similar to $R$(\chiBQ/\chitQ). However, experimental proxies of $R$(\chiBQ/\chiQS) incorporating kinematic cuts exhibit similar $eB$-sensitivities compared to proxies without cuts and the HRG results. Notably, $R^{cut1}_{\rm ALICE}$, $R^{cut}_{\rm STAR}$, and the proxy without any kinematic cuts correspond to about 79\%, 75\%, and 84\% of the lattice QCD results at the strongest $eB$, respectively. 
Specifically, at $eB \sim 5\,M_\pi^2$, the enhancements of $R$(\chiBQ/\chiQS) to its value at zero magnetic fields approximated by the proxies  $R^{cut1}_{\rm ALICE}$, $R^{cut2}_{\rm ALICE}$ and $R^{cut}_{\rm STAR}$, reach approximately 25\%. In contrast, for $R$(\chiBQ/\chitQ), a similar level of enhancement is achieved only by $R^{cut1}_{\rm ALICE}$ at the highest $eB$ values shown in the top left panel of~\autoref{fig:double_rcp_Tpc}. Remarkably, at the largest magnetic fields investigated, the experimental proxies of $R$(\chiBQ/\chiQS) for both ALICE (set1) and STAR kinematic cuts can reach approximately $1.6-1.75$. These findings suggest that $R$(\chiBQ/\chiQS) may serve as an even more effective probe than $R$(\chiBQ/\chitQ).


The bottom left panel of~\autoref{fig:double_rcp_Tpc} presents results for $R$(\chiBQ/\chitB). Unlike $R$(\chiBQ/\chitQ) and $R$(\chiBQ/\chiQS), which exhibit strong $eB$-sensitivity with enhancements reaching $\sim$2 and $\sim$2.25, respectively, $R$(\chiBQ/\chitB) shows a significantly weaker response. Its deviation from the $eB=0$ baseline remains modest, plateauing at $\sim$1.5 for the largest $eB$. While the HRG model underestimates lattice QCD results by $\sim$10\%, the experimental proxies---both with and without kinematic cuts---display negligible differences and only trace a mild increase, rising to $\sim$1.01 at $eB \sim 5\,M_\pi^2$ and $\sim$1.2 at the strongest fields. This weaker dependence on $eB$ is consistent with experimental observations from STAR~\cite{STAR:2019ans}, where only minimal centrality dependence is reported.

The bottom right panel of ~\autoref{fig:double_rcp_Tpc} displays the double ratio $R$(\chitB/\chitQ) along the transition line. Among all studied ratios, this exhibits the smallest sensitivity to $eB$, reaching only about 1.2 at the largest $eB$. Notably, unlike the other cases, the proxy without kinematic cuts slightly overshoots the lattice QCD results for $eB \gtrsim 0.08$ GeV$^2$, reaching approximately 1.4 at the highest fields. Meanwhile, the proxies incorporating kinematic cuts show a trend similar to that observed for $R$(\chiBQ/\chitQ) in the top left panel of ~\autoref{fig:double_rcp_Tpc}, but with larger deviations from unity in the intermediate region of $0.02 \lesssim eB \lesssim 0.09$ GeV$^2$. At the largest $eB$, $R^{cut1}_{\rm ALICE}$ reaches about 1.1, whereas $R^{cut2}_{\rm ALICE}$ and $R^{\rm cut}_{\rm STAR}$ drops to approximately 0.75 and 0.85, respectively.

\section{In extremely strong magnetic fields}
\label{sec:strong-eB}

In previous studies, we primarily investigated second-order fluctuations and correlations of conserved charges in the regime of weak magnetic fields ($eB \lesssim 0.15~\mathrm{GeV}^2$), where the pseudocritical transition temperature $T_{pc}(eB)$ remains nearly constant. Here, we extend this analysis to the strong magnetic field regime ($eB \lesssim 0.8~\mathrm{GeV}^2$), revealing striking contrasts in the system’s response. Fig.~\ref{fig:suscp-vseB_strong_T145} presents new lattice results for second-order fluctuations and correlations at $T = 145~\mathrm{MeV}$, building on earlier findings from Figs.~\ref{fig:charge-suscp-T145} and~\ref{fig:neutral-suscp-T145}. To access stronger fields, we expanded our dataset with simulations at larger magnetic fluxes ($N_b = 12, 16, 24, 32$) for both $N_\tau = 8$ and $N_\tau = 12$, achieving magnetic strengths up to $eB \simeq  0.8~\mathrm{GeV}^2 \sim 45\,M_\pi^2$.

\begin{figure*}[!htp]    
\centering
\includegraphics[width=0.32\textwidth]{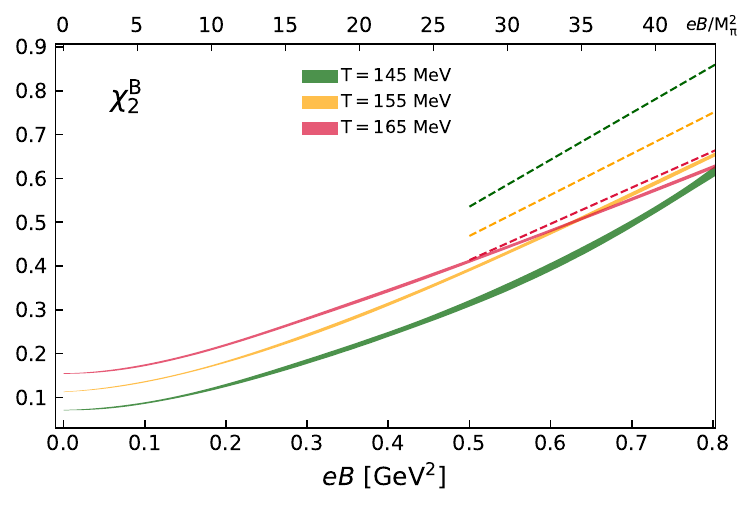}
\includegraphics[width=0.32\textwidth]{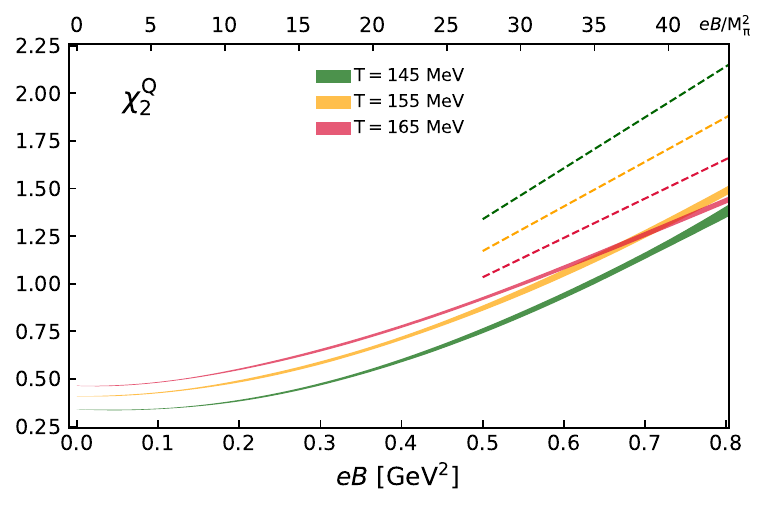}
\includegraphics[width=0.32\textwidth]{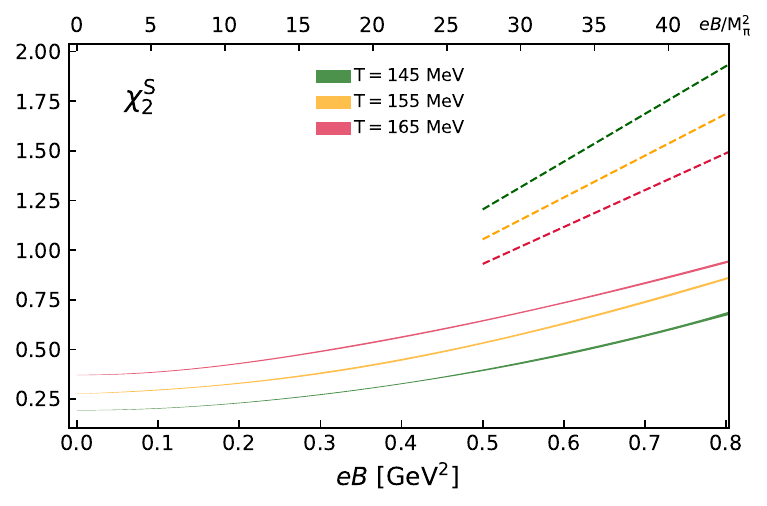}
\includegraphics[width=0.32\textwidth]{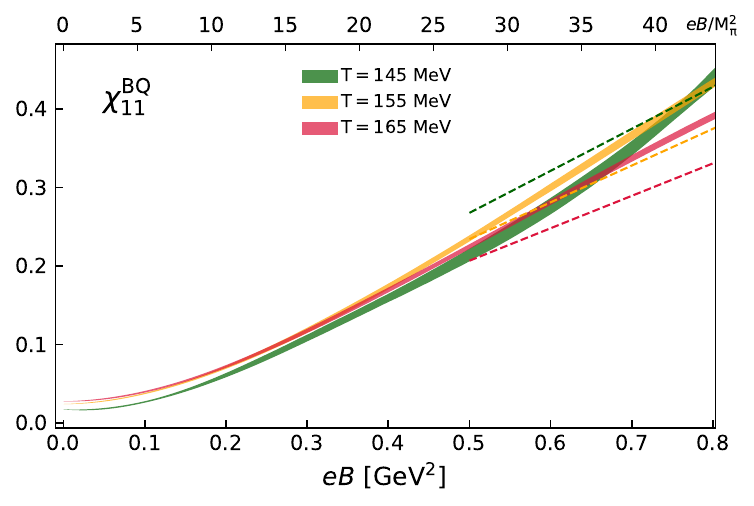}
\includegraphics[width=0.32\textwidth]{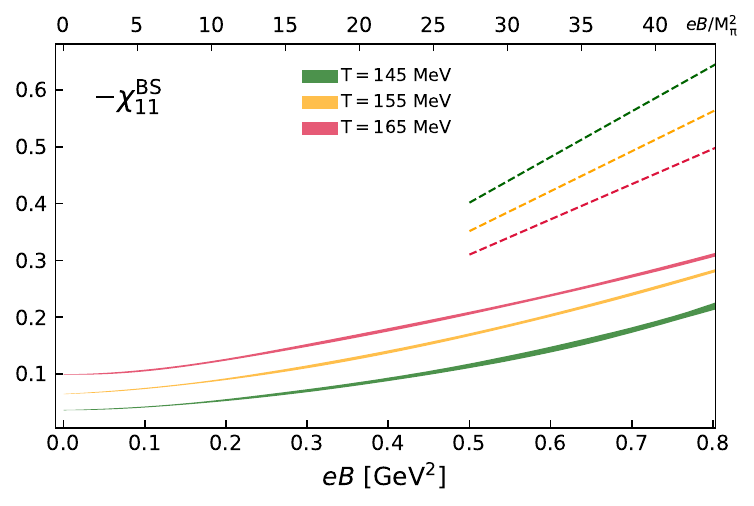}
\includegraphics[width=0.32\textwidth]{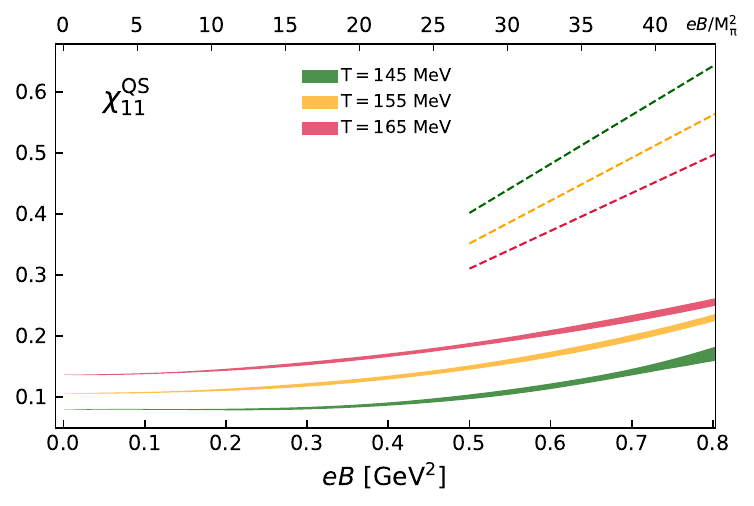}
\caption{$eB$-dependence for continuum estimates of leading-order conserved charges susceptibilities, $\chi^{\rm B}_{2}$, $\chi^{\rm Q}_{2}$, $\chi^{\rm BQ}_{11}$,$\chi^{\rm S}_{2}$, $-\chi^{\rm BS}_{11}$ and $\chi^{\rm QS}_{11}$ (top left to bottom right), at three fixed temperatures, $T=145~{\rm MeV}$ (red), $T=155~{\rm MeV}$ (blue) and $~T=165~{\rm MeV}$ (gold),  around $T_{pc}(eB=0)$ up to strong-$eB$ regime. The dashed lines represent the results from the magnetized ideal gas limit with $\sqrt{eB}/T\rightarrow \infty$ listed in \autoref{tab:free_limit}.}
\label{fig:suscp-vseB_strong}
\end{figure*}

A distinct difference between weak and strong magnetic fields emerges: observables previously insensitive at weaker fields, such as \chitQ, \chiQS~and \chitS, now exhibit pronounced growth, signifying that magnetic interactions become dominant. Among all observables, the baryon-charge correlation \chiBQ~remains the most sensitive, exhibiting approximately a 24-fold enhancement at $eB\sim 0.8$ GeV$^2$, markedly exceeding the baryon fluctuations \chitB~(approximately 8-fold increase), baryon-strangeness correlations \chiBS~(approximately 6-fold increase) and electric charge fluctuations \chitQ~(approximately 4-fold increase).

Interestingly, the charged-strangeness correlation \chiQS~shows the weakest sensitivity, even trailing behind the strangeness fluctuation \chitS~and other observables. Explicitly, the observed hierarchy in magnetic field sensitivity is: \chiBQ $>$  \chitB $>$  \chiBS $>$ \chitQ $>$  \chitS $>$ \chiQS. This hierarchy indicates competing mechanisms: partial cancellations from charged resonance contributions may occur, while nonperturbative effects associated with heavier quarks, especially strange quarks, become more pronounced in strong magnetic fields. The enhancement of strangeness-related observables, when compared to the zero-field case, is at least a factor of 2 at the strongest magnetic fields. This enhancement further highlights the intricate interplay between magnetic fields and quark mass scales, providing valuable insights into the reorganization of QCD matter under extreme conditions.

As previously noted, the HRG model incorporating Landau levels in~\autoref{eq:HRG_energy_level}, provides a reasonable description for the lattice data in the weak-$eB$ regime. However, beyond this range, HRG exhibits discrepancies relative to lattice data. The discrepancies become evident around $eB \sim 10\, M^2_{\pi}$, where charged susceptibilities---$\chi^{\rm BQ}_{11}$, $\chi^{\rm Q}_{2}$, and $\chi^{\rm QS}_{11}$---show a sharp rise and considerably overshoot lattice estimates. In contrast, the baryon fluctuation $\chi^{\rm B}_{2}$ undergoes comparably delayed but similarly pronounced escalation---presumably owing to the influence of magnetically inert neutral hadrons. On the other hand, strangeness-related susceptibilities, $\chi^{\rm S}_{2}$ and $\chi^{\rm BS}_{11}$, dominated by neutral and comparatively heavier strangeness resonances, undershoot the lattice estimates, indicating that the onset of escalation seen in the charged sector is potentially delayed until even stronger fields. The steep increase in charged susceptibilities arises predominantly from the sharply decreasing energies of the lowest Landau levels for higher-spin charged resonances. Notably, doubly charged $\Delta(1232)$, whose energies drop precipitously at $eB \sim 0.2~{\rm GeV}^2$  dominate $\chi^{\rm BQ}_{11}$, contributing for over 50\% of its total value at $eB\sim 0.2~{\rm GeV}^2$. This indicates that the HRG model becomes inapplicable and fails to accurately capture QCD physics even at $T = 145~{\rm MeV}$. A further increase in magnetic strength to $eB \sim 0.29~{\rm GeV}^2$ renders the dispersion relation complex, ultimately leading to the complete breakdown of the HRG framework.

\begin{figure*}[!htp]    
\centering
\includegraphics[width=0.32\textwidth]{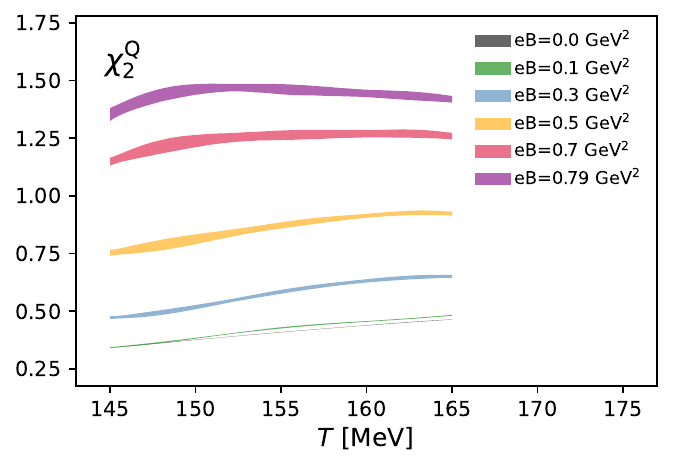}
\includegraphics[width=0.32\textwidth]{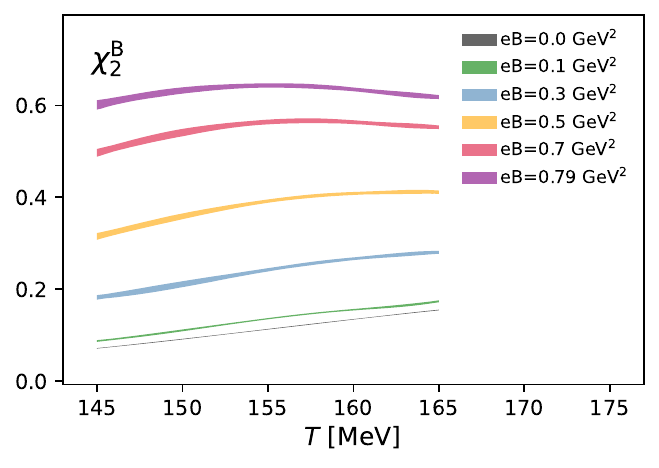}
\includegraphics[width=0.32\textwidth]{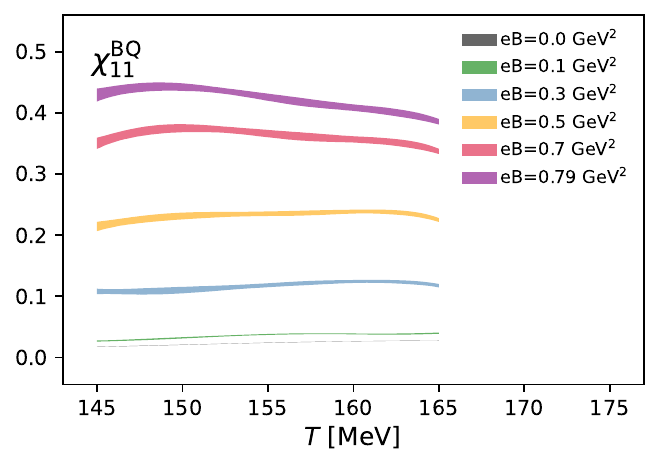}
\includegraphics[width=0.32\textwidth]{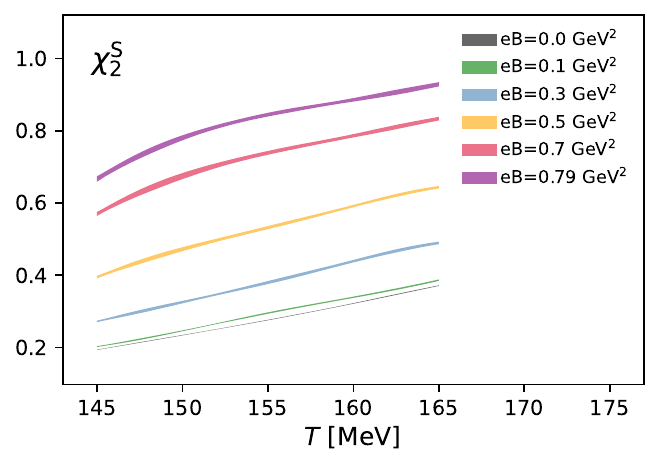}
\includegraphics[width=0.32\textwidth]{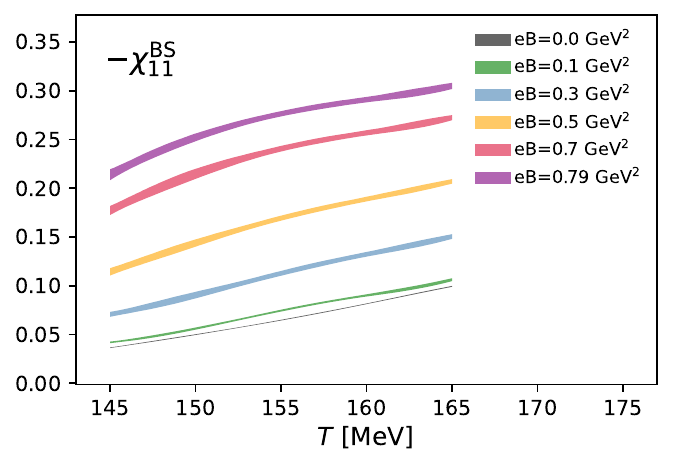}
\includegraphics[width=0.32\textwidth]{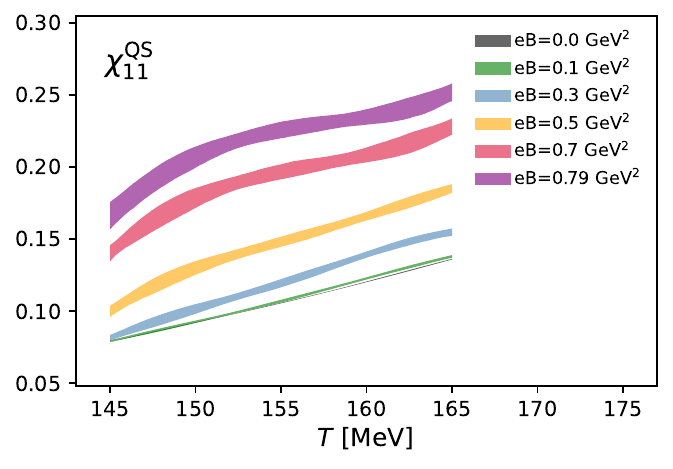}
\caption{$T$ dependence for continuum estimates of leading-order conserved charges susceptibilities, $\chi^{\rm Q}_{2}$, $\chi^{\rm B}_{2}$, $\chi^{\rm BQ}_{11}$,$\chi^{\rm S}_{2}$, $-\chi^{\rm BS}_{11}$ and $\chi^{\rm QS}_{11}$ (top left to bottom right), for fixed strength $eB$ spanning vanishing-, weak- and strong-$eB$ regime.}
\label{fig:suscp-vsT_strong}
\end{figure*}

We now extend our discussion from $T = 145~{\rm MeV}$ to higher temperatures, aiming to elucidate further the interplay between thermal and magnetic effects in the strong-$eB$ regime of QCD. ~\autoref{fig:suscp-vseB_strong} shows the lattice continuum estimates of second-order susceptibilities and correlations as a function $eB$ at fixed temperatures, indicated by colored bands. To clearly depict the interplay between magnetic and thermal effects, we select three representative equally spaced temperature intervals around $T_{pc}(0)$: low-$T = 145~{\rm MeV}$, near $T_{pc}(eB=0)$ with $T = 155~{\rm MeV}$, and high-$T = 165~{\rm MeV}$. As anticipated, all susceptibilities exhibit continuous enhancement with increasing $eB$ across these fixed temperatures, consistent with expectations from the weak-$eB$ regime. However, 
in contrast to the weak-$eB$ regime, a notable new phenomenon emerges at stronger magnetic fields: distinct crossings among temperature bands are observed for $\chi^{\rm B}_{2}$ (top left), $\chi^{\rm Q}_{2}$ (top middle) and $\chi^{\rm BQ}_{11}$ (bottom left). 
These crossings reflect a breakdown of the monotonic temperature hierarchy seen in weak fields, where $\chi^{\rm B}_{2}$, $\chi^{\rm Q}_{2}$ and $\chi^{\rm BQ}_{11}$ rise uniformly with $T$. This nonmonotonicity signals profound modifications to the thermodynamics of charged resonances under extreme magnetic fields, which amplify their contributions to fluctuations and correlations involving baryon number $\rm B$ and electric charge $\rm Q$. 
In sharp contrast, strangeness-related susceptibilities, including strangeness fluctuation $\chi^{\rm S}_{2}$ and the correlation $\chi^{\rm BS}_{11}, \chi^{\rm QS}_{11}$, exhibit no such crossings in the current $eB$-window. 
This observation underscores distinct dynamical effects for different conserved charges and further motivates detailed comparisons with the ideal gas (free) limits, which we explore next.

In regions of extremely strong magnetic fields, the magnetized ideal gas model provides a useful reference for our lattice QCD results. The colored dashed lines in Fig.~\ref{fig:suscp-vseB_strong} denote the free limit (with $\sqrt{eB}/T \to \infty$) of the magnetized ideal gas model discussed in Sec.~\ref{subsec:igl}. In the weak-$eB$ regime, all susceptibilities increase monotonically with temperature, establishing a temperature hierarchy where higher temperatures correspond to a larger magnitude of susceptibility. However, in the strong-$eB$ regime, susceptibilities involving baryon number and electric charges, $\chi^{\rm B}_2$, $\chi^{\rm Q}_2$, and $\chi^{\rm BQ}_{11}$, exhibit crossings between fixed temperature bands, signaling a reordering of this hierarchy. As the magnetic field is increased further, the lattice results appear to approach the ideal gas limit, where this hierarchy is ultimately reversed. This reversal is partly driven by the dominant contribution of the highly degenerate lowest Landau level, which is enhanced at lower temperatures in the ideal gas due to reduced thermal excitations and the anticipated dimensional reduction \cite{Gusynin:1994re,Gusynin:1995nb}. Notably, for the strangeness sector, no crossings or reordering occur for $eB \lesssim 0.8\,\mathrm{GeV}^2$, and these results remain furthest from the free limit. It is crucial to note that, fundamentally, these nontrivial crossings and reordering are manifestations of significant changes in the temperature dependence of the susceptibilities.

Fig.~\ref{fig:suscp-vsT_strong} illustrates the temperature dependence of the lattice QCD continuum estimates for all leading-order susceptibilities. Unlike in Fig.~\ref{fig:suscp-vseB_strong}, where the colored bands correspond to fixed temperatures, here they represent continuum estimates at constant magnetic field strengths, $eB$. This allows us to systematically examine the $T$-dependence across different fixed-$eB$ bands in both the weak- and strong-$eB$ regimes. Specifically, we consider $eB=0$ for the vanishing magnetic field case, $eB=0.1~{\rm GeV}^2$ in the weak-$eB$ regime, and $eB=0.3,~ 0.5,~ 0.7$ and $0.79~{\rm GeV}^2$ in the strong-$eB$ regime.  

In the absence of a magnetic field, all second-order susceptibilities exhibit the expected monotonic increase with temperature, driven by thermal agitation. However, upon introducing a weak magnetic field, an enhanced growth emerges near $T_{pc}(eB) \sim T_{pc}(0)$, indicating a nontrivial interplay between thermal and magnetic effects. More strikingly, as the field strength enters the strong-$eB$ regime, the key susceptibilities---$\chi^{\rm Q}_{2}$ (top left), $\chi^{\rm B}_{2}$ (top middle), and $\chi^{\rm BQ}_{11}$ (top right)---undergo a rapid escalation and develop nonmonotonic features. This effect is particularly pronounced for $\chi^{\rm B}_{2}$, which begins to exhibit a mild peak structure at $eB\gtrsim 0.7~{\rm GeV}^2$.

\begin{figure*}[!htp]    
\centering
\includegraphics[width=0.32\textwidth]{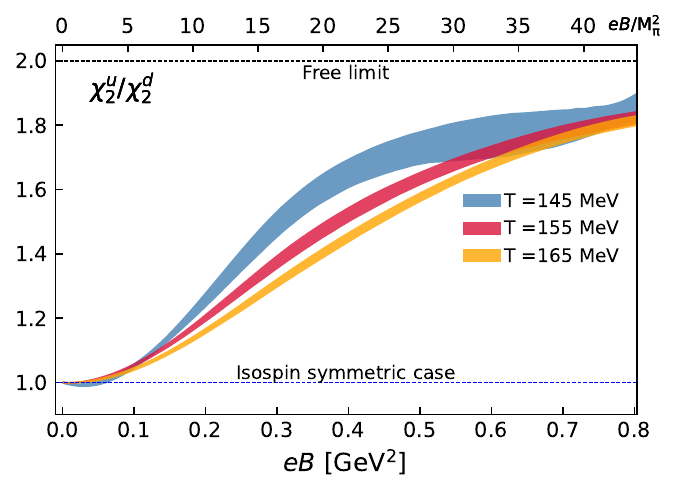}
\includegraphics[width=0.32\textwidth]{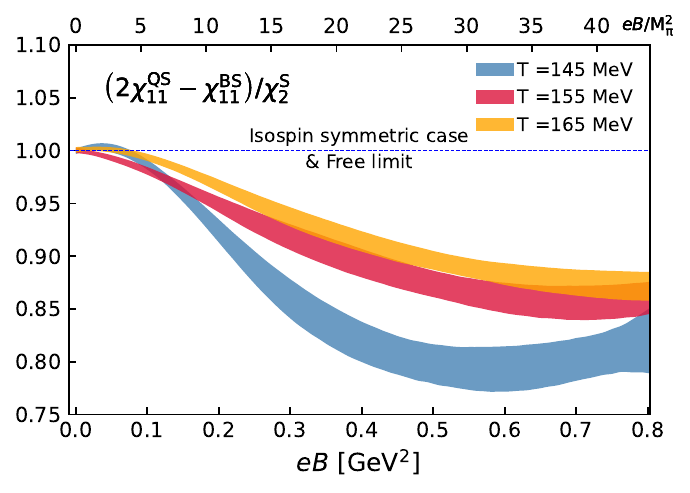}
\includegraphics[width=0.32\textwidth]{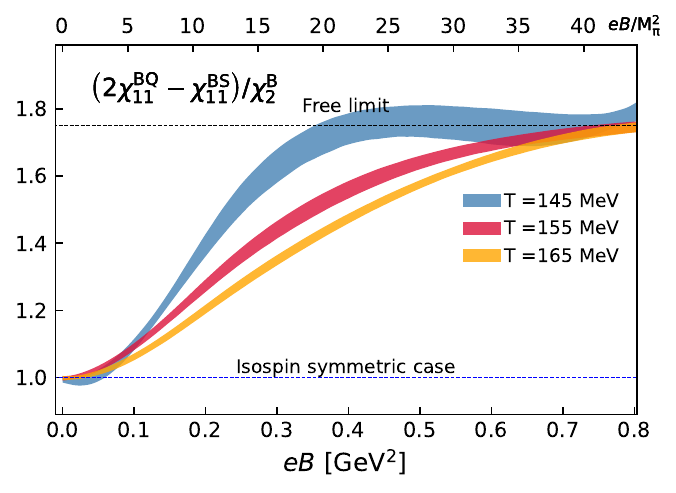}
\caption{Isospin symmetry breaking effects manifested in fundamental quark fluctuations $\chi^u_2 / \chi^d_2$ (left) and physical conserved charges basis $\left(2\chi^{\rm QS}_{11} - \chi^{\rm BS}_{11}\right) / \chi^{\rm S}_{2}$ (middle) and $\left(2\chi^{\rm BQ}_{11} - \chi^{\rm BS}_{11}\right) / \chi^{\rm B}_{2}$ (right). The colored band represents continuum estimates versus $eB$ up to the strong-$eB$ regime at three fixed temperatures, $T=145~{\rm MeV}$ (red), $T=155~{\rm MeV}$ (blue) and $~T=165~{\rm MeV}$ (gold), around $T_{pc}(eB=0)$.}
\label{fig:iso-sym_seB}
\end{figure*}

Notably, the inflection points and peak locations undergo a systematic shift toward lower temperatures, suggesting that the dominant degrees of freedom associated with baryon number and electric charge undergo transitions at reduced temperatures in stronger magnetic fields. In the literature, this downward shift in the transition temperature due to an external magnetic field is commonly referred to as the $T_{pc}$-lowering effect and is in agreement with previous studies on light and strange quark chiral condensates 
\cite{Bali:2011qj,Ding:2022tqn} and Polyakov loop \cite{Bruckmann:2013oba,DElia:2018xwo}.
In contrast, strangeness-related susceptibilities, \chitS~and their correlations \chiBS~and \chiQS~ in ~\autoref{fig:suscp-vsT_strong}, exhibit monotonic temperature dependence without crossings, as already inferred from their $eB$-dependence in~\autoref{fig:suscp-vseB_strong}. However, their temperature dependence becomes notably steeper at strong magnetic fields, particularly at $eB \simeq 0.8~{\rm GeV}^2$ with a steeper slope emerging in the lower temperature region. 

Before concluding our discussion of conserved charges susceptibilities in extremely strong magnetic fields, we note that the observed nonmonotonic features, emerging mild peak structures, and the lowering of $T_{pc}$ are qualitatively consistent with earlier investigations using heavier-than-physical pion masses~\cite{Ding:2021cwv} and with a recent pseudoscalar meson study at the physical pion mass~\cite{Ding:2025pbu}. Furthermore, earlier results also suggest that substantially more prominent peaks and a more pronounced $T_{pc}$-lowering may emerge at even stronger magnetic fields.

\section{Isospin symmetry breaking}
\label{sec:isospin}

In our $(2+1)$-flavor QCD calculations, we treat the light up and down quarks as degenerate isospin partners ($m_u = m_d$), as is commonly done in lattice QCD studies at nonzero temperatures. At $eB=0$, this treatment preserves isospin symmetry. However, the presence of a magnetic field explicitly breaks the symmetry due to the distinct electric charges of the up and down quarks. Although isospin symmetry is intrinsically violated in the physical scenario where $m_u \neq m_d$, the effects are relatively subtle compared to the significantly stronger and more pronounced symmetry breaking induced by the magnetic field. Such breaking can be directly investigated through the ratio of second-order fluctuations of up and down quark numbers, $\chi^u_2 / \chi^d_2$. In the absence of magnetic fields, the ratio $\chi^u_2 / \chi^d_2$ is expected to remain unity across all temperatures, while in the magnetized ideal gas limit (with $\sqrt{eB}/T \to \infty$), it asymptotically approaches 2, as discussed in Sec. \ref{sec:fluc_HRG_ideal}.

In the left panel of Fig. \ref{fig:iso-sym_seB}, we present the $eB$-dependence for the continuum estimates of the ratio $\chi^u_2 / \chi^d_2$ at three fixed temperatures around $T_{pc}(eB)$; lower at $T = 145~{\rm MeV}$, near $T_{pc}(0)$ at $T = 155~{\rm MeV}$, and higher at $T = 165~{\rm MeV}$. The lattice results show consistency with our theoretical expectations, as $\chi^u_2 / \chi^d_2$ increases monotonically from unity and eventually approaches saturation toward the free limit near the end of the strong-$eB$ regime. Notably, in the strong-$eB$ regime, particularly $eB\gtrsim 0.15~{\rm GeV}^2 (\sim 8\,M^2_{\pi})$, the low-temperature $T= 145~{\rm MeV}$ band exhibits heightened sensitivity to isospin symmetry breaking. However, toward the end of the strong-$eB$ regime, the higher $T=165~{\rm MeV}$ band eventually catches up, with all fixed temperature bands agreeing within uncertainties and approaching saturation to the free limit.

Note that similar clear hints of isospin symmetry breaking are also evident in chiral condensates ~\cite{Ding:2025pbu,Ding:2020hxw,Bali:2012zg,DElia:2011koc}, yet neither quark fluctuations nor condensates-based effects can be measured directly in experiments. In principle, conserved charges provide an avenue to probe these quark fluctuations; however, precise measurements of $\chi^{u}_2$ and $\chi^{d}_2$ through conserved charge
\begin{align}
    \label{eqn:iso-ud-BQS}
   \chi^{u}_{2} &= \chi^{\rm B}_{2} + \chi^{\rm Q}_{2} + 2\chi^{\rm BQ}_{11}  \,,\\
    \chi^{d}_{2} &= 4\chi^{\rm B}_{2} + \chi^{\rm Q}_{2} + \chi^{\rm S}_{2} - 4\chi^{\rm BQ}_{11} - 2\chi^{\rm QS}_{11} +4\chi^{\rm BS}_{11}\,,
\end{align}
pose significant complexity challenges in experiments. Consequently, exploring isospin symmetry through appropriate, experimentally accessible combinations of conserved charge susceptibilities remains a promising direction for further investigation. In vanishing magnetic fields, it is well established that the degeneracy between $u$- and $d$-quark fluctuations and their respective correlations with the strange quark ($\chi^{us}_{11} = \chi^{ds}_{11}$) reduces the six nominal leading-order cumulants in conserved charge basis to four independent ones \cite{Bazavov:2017dus,Bollweg:2021vqf,Ding:2021cwv}, constrained by the following two relations:
\begin{align}
    \label{eqn:iso-symm-QS}
   2\chi^{\rm QS}_{11} - \chi^{\rm BS}_{11} &= \chi^{\rm S}_{2} \,,\\
    \label{eqn:iso-symm-BQ}
    2\chi^{\rm BQ}_{11} - \chi^{\rm BS}_{11} &=\chi^{\rm B}_{2}\,.
\end{align}
These constraints can serve as more pragmatic probes of isospin symmetry breaking by redefining observables as $\left(2\chi^{\rm QS}_{11} - \chi^{\rm BS}_{11}\right) / \chi^{\rm S}_{2}$ and $\left(2\chi^{\rm BQ}_{11} - \chi^{\rm BS}_{11}\right) / \chi^{\rm B}_{2}$, which may also be explored in experimental searches.

First, let us examine the observable $ \left(2\chi^{\rm QS}_{11} - \chi^{\rm BS}_{11}\right) / \chi^{\rm S}_{2} $. Interestingly, this ratio remains unity not only at the vanishing magnetic field but also in the magnetized ideal gas, owning to Eq. \eqref{eqn:iso-symm-QS} being satisfied by virtue of the absence of quark number correlations. This behavior suggests that either isospin symmetry is preserved at $ eB \neq 0 $ or a nonmonotonic trend would emerge. The middle panel of Fig. \ref{fig:iso-sym_seB} shows $ \left(2\chi^{\rm QS}_{11} - \chi^{\rm BS}_{11}\right) / \chi^{\rm S}_{2} $ as a function of $eB$ for three fixed temperatures around $T_{pc}(eB=0)$, analogous to $\chi^u_2 / \chi^d_2$ plot. As magnetic fields are introduced, apparently the ratio begins to deviate from unity, and while this deviation initially grows, theoretically a turning point is anticipated in the stronger $eB$ regime, beyond which the ratio should converge toward the magnetized free limit. Interestingly, we again observe that low-temperature physics appears more sensitive to isospin-breaking effects. Nonetheless, the deviations from unity remain relatively small, at most $20\%$ at $eB\sim0.6~{\rm GeV}^2~(\sim35\, M^2_{\pi}) $ and are expected to diminish further beyond the turning point.

Next, we consider $ \left(2\chi^{\rm BQ}_{11} - \chi^{\rm BS}_{11}\right) / \chi^{\rm B}_{2} $. Unlike the previously discussed probe in Eq. \eqref{eqn:iso-symm-QS}, the relation in Eq. \eqref{eqn:iso-symm-BQ} does not hold for the magnetized ideal gas. Using expressions outlined in Table \ref{tab:free_limit}, the ideal gas limit (with $\sqrt{eB}/T\rightarrow \infty$) can be straightforwardly computed,
\begin{equation}
    \lim_{\sqrt{eB}/T \to \infty}\left(2\chi^{\rm BQ}_{11} - \chi^{\rm BS}_{11}\right) /\chi^{\rm B}_{2} = 7/4.
\end{equation}
The ratio $\left(2\chi^{\rm BQ}_{11} - \chi^{\rm BS}_{11}\right)/\chi^{\rm B}_{2} $ is expected to grow from unity toward $7/4$ as $ eB $ increases. The right panel of Fig. \ref{fig:iso-sym_seB} illustrates this behavior. The low-temperature $T=145~{\rm MeV}$ band reaches the ideal gas limit rather sharply at $eB \sim 0.35~{\rm GeV}^2~(\sim20 \,M^2_{\pi}) $ and saturates, whereas the high temperature $T= 165~{\rm MeV} $ band approaches the limit more steadily, extending up to $eB \sim 0.8~{\rm GeV}^2 ~(\sim 45\, M^2_{\pi}) $. Qualitatively similar trends were observed in earlier studies~\cite{Ding:2021cwv}.

From an experimental perspective, although $\left(2\chi^{\rm BQ}_{11} - \chi^{\rm BS}_{11}\right)/\chi^{\rm B}_{2}$ may appear relatively feasible, it is important to point out that up to $eB\lesssim 8\,M_{\pi}^2$, the enhancements of both $\left(2\chi^{\rm BQ}_{11} - \chi^{\rm BS}_{11}\right)/\chi^{\rm B}_{2}$ and $ \left(2\chi^{\rm QS}_{11} - \chi^{\rm BS}_{11}\right) / \chi^{\rm S}_{2} $ are rather marginal---at most $\sim 25\%$ and $\sim 5\%$, respectively---especially when compared to the proposed $\chi^{\rm BQ}_{11}$ magnetometer~\cite{Ding:2023bft}.

\section{Conclusion}
\label{sec:summary}

In this work, we conducted a systematic lattice QCD investigation of second-order fluctuations of and correlations among conserved charges (baryon number, electric charge, and strangeness) in the presence of strong magnetic fields. Utilizing the state-of-the-art lattice QCD simulations with physical pion masses and performing continuum estimates, we explored magnetic field strengths extending up to unprecedented levels ($eB \simeq 0.8~\text{GeV}^2$). 
Our analysis establishes a clear hierarchy of sensitivity to magnetic fields across observables, with the baryon-electric charge correlation \chiBQ~emerging as a particularly powerful magnetometer for probing magnetized QCD matter.

In the regime of relatively weak magnetic fields $eB \lesssim 8\,M_{\pi}^2$, while the HRG model does not fully describe susceptibilities near $T_{pc}(eB)$, it remains indispensable for connecting lattice QCD results to experimental observables. By leveraging its capability to map conserved charge fluctuations to detectable final-state particles---such as protons for baryon number and kaons for strangeness---the HRG provides a practical framework for constructing proxies. These proxies, incorporating kinematic cuts emulating STAR and ALICE detector acceptances, retain approximately 80\% of the magnetic sensitivity predicted by lattice QCD (~\autoref{fig:rcp_Tpc}). This high sensitivity retention arises from the HRG’s ability to approximate relative \(eB\)-dependent trends and the suppression of volume effects in ratio observables. Optimized double ratios like $R(\chi_{11}^{\rm BQ}/\chi_{2}^{\rm Q})$ and $R(\chi_{11}^{\rm BQ}/\chi_{11}^{\rm QS})$ further enhance experimental feasibility, exhibiting up to 25\% enhancements at $eB \sim 0.15~{\rm GeV}^2$ even after kinematic cuts (~\autoref{fig:double_rcp_Tpc}). This underscores their potential to isolate magnetic field signatures in heavy-ion collisions, where direct measurements of $eB$ remain challenging. Thus, by bridging theoretical predictions with detector-level analyses, the HRG-based proxies offer a pathway for probing magnetic effects in QCD matter through accessible fluctuation observables.

As the magnetic field strength increases beyond $eB \sim 8\,M_\pi^2$, substantial deviations emerge between lattice QCD results and HRG model predictions. These discrepancies highlight nontrivial modifications of hadronic degrees of freedom---likely driven by magnetic field-induced changes in hadron spectra---that cannot be captured by conventional HRG treatments. Notably, we observed nonmonotonic temperature dependence in \chitB, \chitQ~and \chiBQ~at $eB \gtrsim 0.5~\text{GeV}^2$, signaling a complex interplay between thermal and magnetic energy scales. These findings underscore the necessity of first-principles lattice QCD studies in regimes where magnetic fields dominate over thermal effects.

Our results also reveal pronounced isospin-breaking effects in fluctuation observables under strong magnetic fields (cf.~\autoref{fig:iso-sym_seB}). The magnetic environment induces substantial differences between up- and down-quark susceptibilities, reflecting the distinct electric charges and magnetic responses of light quarks. This breaking of isospin symmetry manifests in measurable deviations of charge-dependent correlations, offering a novel pathway to probe quark-level dynamics in magnetized QCD matter.

The baryon-electric charge correlation \chiBQ~ stands out as a uniquely sensitive probe of magnetic fields in QCD matter, bridging theoretical predictions and experimental feasibility. Our work provides QCD benchmarks for interpreting current and future heavy-ion collision data, particularly in disentangling the interplay of thermal and magnetic effects. The breakdown of HRG descriptions at strong fields emphasizes the need for further studies of magnetically modified hadron properties, while the constructed proxies establish a concrete framework for experimental validation. These advances pave the way for a deeper exploration of QCD under extreme magnetic conditions, with implications for understanding the quark-gluon plasma and the role of magnetic fields in relativistic nuclear collisions.

We also note that a new set of kinematic cuts (ALICE set2), introduced in a recent experimental analysis~\cite{ALICE:2025mkk} which appeared on the same day as our initial submission, has been incorporated into the current version of our study. While the experimental work focuses on the double ratio \chiBQ/\chitQ, we find that the ratio \chiBQ/\chiQS, though not yet reported experimentally, provides a particularly robust theoretical signature. This observable shows enhanced sensitivity to the presence of a magnetic field and exhibits reduced dependence on specific kinematic cuts, making it a promising complementary probe in future analyses.

\section*{Acknowledgements}
We thank Xiaofeng Luo, Swagato Mukherjee, and Nu
Xu for useful discussions. This research was partially funded by the National Natural Science Foundation of China under Grants No. 12293060, No. 12293064; No. 12325508, along with support from the National Key Research and Development Program of China under Contract No. 2022YFA1604900. Computational resources for the numerical simulations were provided by the GPU cluster at the Nuclear Science Computing Center, Central China Normal University (NSC$^3$), and the Wuhan Supercomputing Center.

\bibliographystyle{JHEP.bst}
\bibliography{refs.bib}

\begin{widetext}
\appendix

\section{Details on simulation parameters}
\label{app:stat}

\begin{table}[!htp]
    \centering
        \begin{tabular}{*{16}{c}}
        \hline \hline
        \multirow{2}*{$N_\sigma^3\times N_\tau$}  
        &	\multirow{2}*{$T$ [MeV]}  
        &	\multirow{2}*{$\beta$}	 	
        &   \multirow{2}*{$am_l$} 
        &   \multirow{2}*{$am_s$} 
        &   \multicolumn{9}{c}{\# conf.}
        &   \multirow{2}*{$N_{\rm v1}$} 
        &   \multirow{2}*{$N_{\rm v2}$} 
        \\
        \cline{6-14} 
        &         &         &        &        & $N_b=1$ & $N_b=2$ & $N_b=3$ & $N_b=4$ & $N_b=6$& $N_b=12$ & $N_b=16$ & $N_b=24$ & $N_b=32$ &   &   \\
        \hline 
                        &144.95& 6.315 & 0.00281 & 0.0759& 50831 & 53826 & 49670 & 57380 & 57416 & 53624 & 49863 & 22580 & 20777 & 603 & 603  \\
                        &151.00& 6.354 & 0.00270 & 0.0728& 50283 & 53229 & 58907 & 55797 & 58079 & 55395 & 44513 & 14884 & 14872 & 603 & 603  \\
        $32^3 \times 8$ &156.78& 6.390 & 0.00257 & 0.0694& 62568 & 56330 & 46670 & 51375 & 48417 & 45768 & 22006 & 21921 & 20335 & 603 & 603  \\
                        &162.25& 6.423 & 0.00248 & 0.0670& 26259 & 24041 & 26619 & 26214 & 32426 & 36968 & 23002 & 23121 & 26022 & 603 & 603  \\
                        &165.98& 6.445 & 0.00241 & 0.0652& 25080 & 20450 & 23103 & 22048 & 24600 & 27122 & 19356 & 23658 & 24592 & 603 & 603  \\
                        \hline
                        &144.97& 6.712 & 0.00181 & 0.0490& 5583  & 5335  & 5326  & 5370  & 5170  & 3380  & 3508  & 3321  & 3318  & 705 & 102  \\
                        &151.09& 6.754 & 0.00173 & 0.0468& 5299  & 5098  & 5130  & 5033  & 5160  & 3519  & 3114  & 3087  & 2656  & 606 & 102  \\
        $48^3 \times 12$&157.13& 6.794 & 0.00167 & 0.0450& 4315  & 4464  & 4161  & 4241  & 4108  & 3080  & 3343  & 3295  & 2847  & 705 & 102  \\
                        &161.94& 6.825 & 0.00161 & 0.0436& 5871  & 2456  & 2820  & 7318  & 4577  & 3039  & 2974  & 2806  & 2634  & 405 & 102  \\
                        &165.91& 6.850 & 0.00157 & 0.0424& 3000  & 3000  & 2560  & 3000  & 2271  & 2454  & 2951  & 2568  & 3231  & 102 & 102  \\
                        \hline
        $64^3 \times 16$&156.92& 7.095 & 0.00124 & 0.0334&   -   &   -   & 3052  &   -   &   -   &   -   &   -   &   -   &   -   &  603  & 102  \\
        \hline \hline	
        \end{tabular}
    \caption{Simulation parameters and statistics on $32^3 \times 8$, $48^3 \times 12$ and $64^3 \times 16$ lattices with light to strange quark mass ratio $m_l/m_s =1/27$. }
    \label{tab:stat}
\end{table}

The parameters and accumulated statistics for lattice QCD simulations on $N_{\tau}$ = 8, 12 and 16 lattices are summarized in~\autoref{tab:stat}. In this table, $N_{\rm v1}$ and $N_{\rm v2}$ denote the number of random noise vectors utilized for computing $D_1=\partial \mathrm{ln~det} M_f/\partial \mu_f|_{\mu_f=0}$ and $D_2=\partial^2 \mathrm{ln~det} M_f/\partial \mu_f^2|_{\mu_f=0}$, respectively, where $M_f$ is the staggered fermion matrix for quark flavor $f$ and $\mu_f$ is the corresponding flavor chemical potential. 

\section{Proxy with kinematic cuts and supplemental plots to~\autoref{fig:double_rcp_Tpc}}
\label{app:proxy_cuts}

\begin{figure*}[!h]   
    \centering           
    \includegraphics[width=0.32\textwidth]{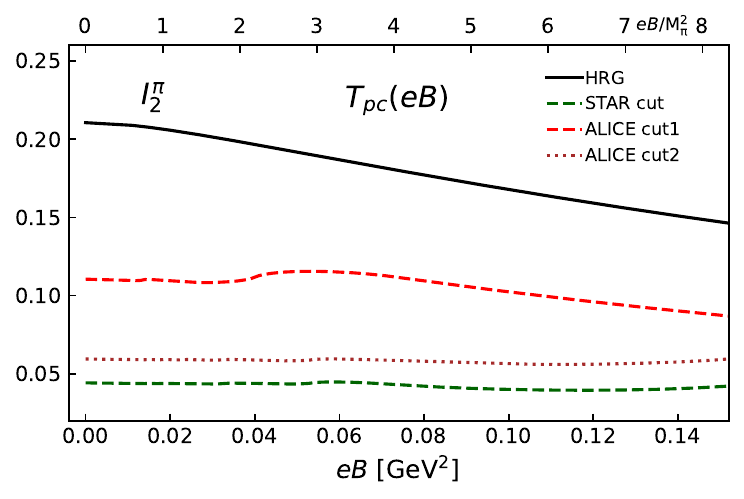}
    \includegraphics[width=0.32\textwidth]{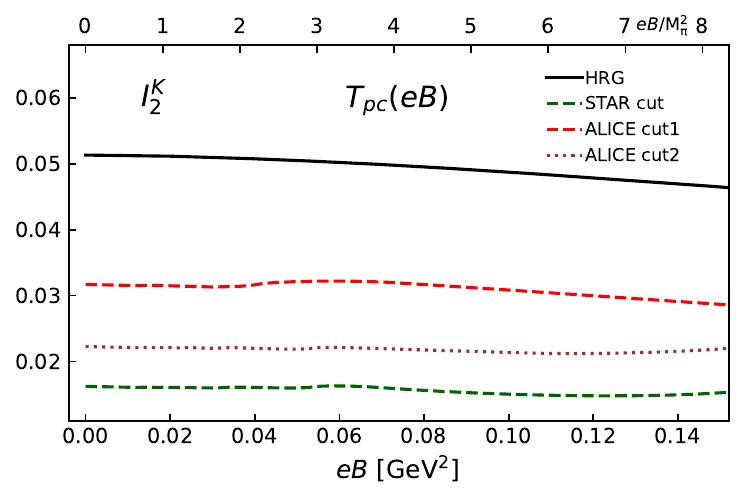}
    \includegraphics[width=0.32\textwidth]{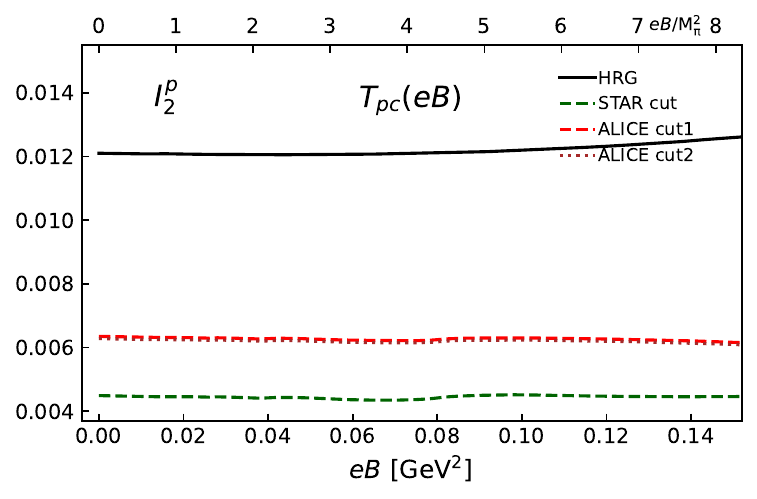}
    \caption{Leading-order phase space integral $I^i_2$ for $\pi, K$ and $p$ along the transition line. Black solid lines represent the full space HRG model results without any cuts. The red and blue dashed lines represent results with STAR and ALICE kinematic cuts, respectively.}
    \label{fig:sup_piKp_kcuts}
\end{figure*}

\begin{figure*}[!htp]  
\includegraphics[width=0.42\textwidth]{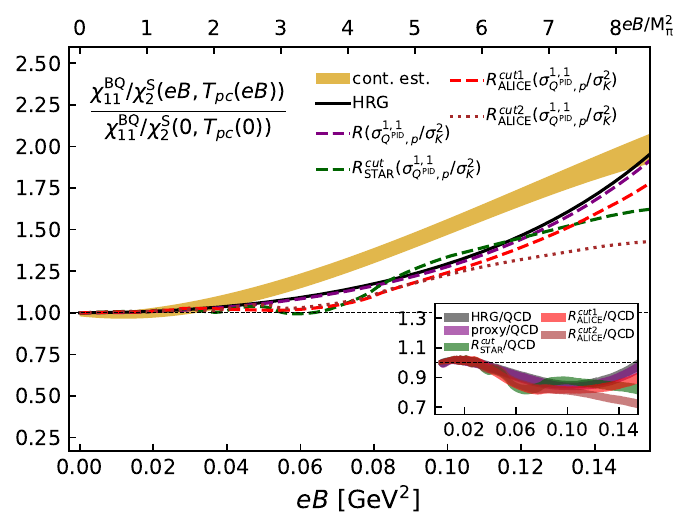}
\includegraphics[width=0.42\textwidth]{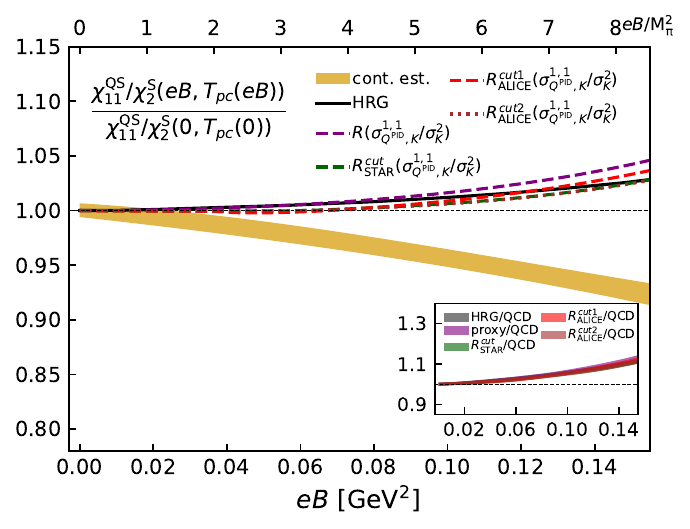}
\includegraphics[width=0.42\textwidth]{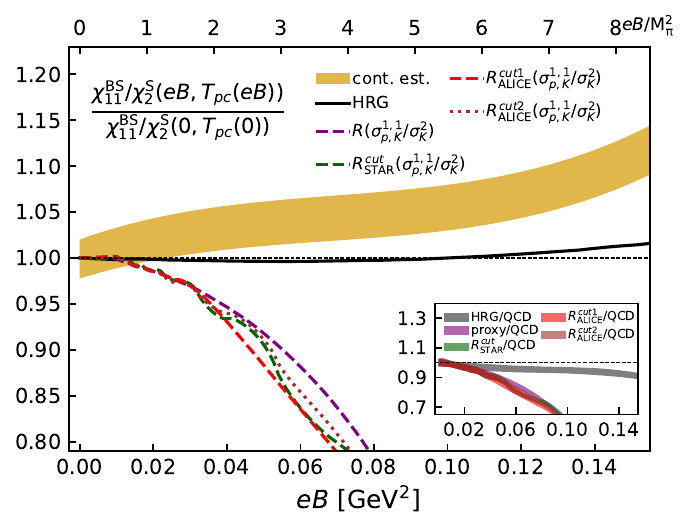}
\includegraphics[width=0.42\textwidth]{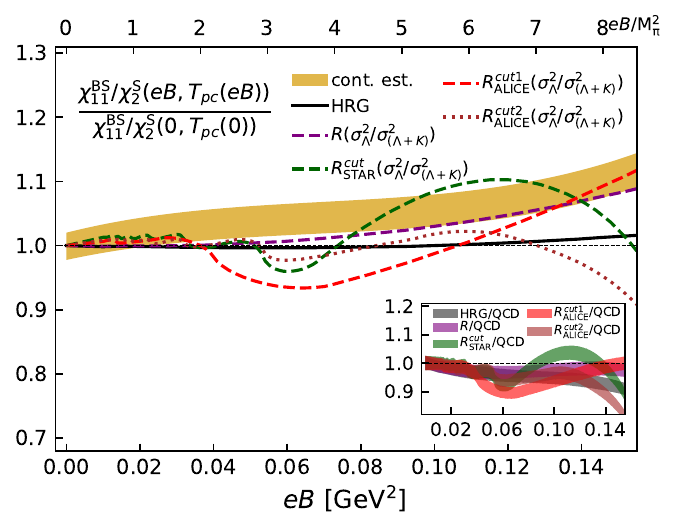}
\caption{Double ratios $R(\chi^{\rm BQ}_{11}/\chi^{\rm S}_{2})$, $R(\chi^{\rm QS}_{11}/\chi^{\rm S}_{2})$ and $R(\chi^{\rm BS}_{11}/\chi^{\rm S}_{2})$  (from top to bottom) along $T=T_{pc}(eB)$. The gold band represents the continuum estimates obtained from lattice QCD simulations. The solid line denotes results calculated from the HRG model, while the dashed lines correspond to experimental proxies: the purple, green, and red dashed lines represent proxies integrated over the full phase space, and those with kinematic cuts applied for the STAR and ALICE detectors, respectively. The insets depict the ratio of the HRG model and experimental proxies to the corresponding lattice QCD continuum estimates.}
\label{fig:app_double_rcp}
\end{figure*}

\autoref{fig:sup_piKp_kcuts} illustrates the leading-order phase space integral $I^i_2$ at $T_{pc}(eB)$ for three particles---$\pi$ (left panel), $K$ (middle panel), and $p$ (right panel)---under three distinct conditions: the full phase space in the Hadron Resonance Gas (HRG) model without kinematic cuts, and with kinematic cuts applied from the STAR and ALICE experiments. The ALICE and STAR kinematic cuts are defined as \autoref{eq:kcuts_ALICE_STAR}.

\autoref{fig:app_double_rcp} presents several double ratio observables related to strangeness, including $R(\chi^{\rm BQ}_{11}/\chi^{\rm S}_{2})$ (top left), $R(\chi^{\rm QS}_{11}/\chi^{\rm S}_{2})$ (top right) and $R(\chi^{\rm BS}_{11}/\chi^{\rm S}_{2})$ (bottom). For $R(\chi^{\rm BQ}_{11}/\chi^{\rm S}_{2})$, it shows similar behavior as $R(\chi^{\rm BQ}_{11}/\chi^{\rm QS}_{11})$ (see~\autoref{fig:double_rcp_Tpc} top right). For $R(\chi^{\rm QS}_{11}/\chi^{\rm S}_{2})$, the proxies show the opposite trend from lattice QCD results. For $R(\chi^{\rm BS}_{11}/\chi^{\rm S}_{2})$, the use of only the $\pi,~K$ and $p$ proxy fails to describe the QCD results over a wide range of magnetic fields (see \autoref{fig:app_double_rcp} bottom left). To address this, we employ the detectable $\Lambda$ particle as a proxy for $\chi^{\rm BS}_{11}$ and the combination of $\Lambda$ and kaon as proxies for $\chi^{\rm S}_{2}$ \cite{Bellwied:2019pxh}. The kinematic cuts on $\Lambda$ applied for STAR are 0.9 GeV/$c$ $< p_T <$ 2 GeV/$c$ and $ y < |0.5| $ \cite{Nonaka:2019fhk}, for ALICE are  1 GeV/$c$ $< p_T <$ 4 GeV/$c$  and $ \eta < |0.5| $ \cite{Umaka:2020ait}. Although the proxy $R(\sigma_\Lambda^2/\sigma^2_{(\Lambda+K)})$  approaches the lower boundary of the QCD results, after applying the kinematic cuts from ALICE and STAR experiments, both  $R^{cut}_{\rm STAR}$, $R^{cut1}_{\rm ALICE}$ and $R^{cut2}_{\rm ALICE}$,  fall within the band of QCD results at $eB\lesssim 2\,M_\pi^2$. As $eB$ grows, these adjusted proxies begin to exhibit behavior similar to that of double ratios for \chiBQ/\chitQ~ (see~\autoref{fig:double_rcp_Tpc} top left). In summary, these results shown in \autoref{fig:app_double_rcp} reinforce the conclusion that observables involving $\chi^{\rm BQ}_{11}$ are particularly well-suited for detecting magnetic fields in heavy-ion collision experiments.

\end{widetext}

\end{document}